\documentstyle[aps,prb,preprint,amssymb]{revtex}
\begin{document}

\draft
\title{Collective Particle Flow through Random Media}
\author{Joe Watson and Daniel S. Fisher}
\address{Physics Department\\ Harvard University}
\date{Submitted to Phys.\ Rev.~B. on September 20, 1995}
\maketitle

\begin{abstract}
  A simple model for the nonlinear collective transport of interacting
  particles in a random medium with strong disorder is introduced and
  analyzed. A finite threshold for the driving force divides the
  behavior into two regimes characterized by the presence or absence
  of a steady-state particle current. Below this threshold, transient
  motion is found in response to an increase in the force, while above
  threshold the flow approaches a steady state with motion only on a
  network of channels which is sparse near threshold. Some of the
  critical behavior near threshold is analyzed via mean field theory,
  and analytic results on the statistics of the moving phase are
  derived. Many of the results should apply, at least qualitatively,
  to the motion of magnetic bubble arrays and to the driven motion of
  vortices in thin film superconductors when the randomness is strong
  enough to destroy the tendencies to lattice order even on short
  length scales. Various history dependent phenomena are also
  discussed.
\end{abstract}

\section{Introduction}
\label{sec:intro}

The collective transport of classical objects through random media
exhibits many interesting features. In the presence of an applied
driving force, steady-state motion is usually only possible, in the
absence of fluctuations, if the driving force is sufficiently large.
The behavior in the vicinity of the threshold force provides a rich
collection of different types of non-equilibrium dynamic critical
phenomena.  These systems can be divided, roughly, into two broad
classes, depending on the relative strength of the interactions of the
objects among themselves and the interactions with the random medium,
the `pinning'.

If the pinning is, in some sense, weak compared to the interactions
among the transported objects then these can form an extended elastic
structure. Transport can then be considered as the motion and
distortion of this structure. Examples of this elastic class are
domain wall motion in magnets%
,\cite{%
koiller:disdomwall,%
ji:percdomain,%
nattermann:dynintdepin,%
narayan:thresh} %
sliding charge density waves%
,\cite{%
sv:phy217,
fisher:threshcdwlett+fisher:scdwdyncrit,%
middleton:pcdwlett+middleton:cdwbelow,%
narayan:scdwepslett+narayan:scdweps,%
narayan:avalrgcdw} 
weakly pinned Abrikosov flux lattices%
,\cite{larkin:pinning,blatter:vortexrmp}
the interfaces of wetting fluids invading porous media%
,\cite{%
lenormand:liqporous,
rubio:selfaff+rubio:sareply,%
horvath:dynvis,
he:roughwet,
nolle:morphint} 
contact lines of spreading droplets on rough surfaces%
,\cite{joanny:contactline,ertas:clcd}
surface growth by random deposition%
,\cite{barabasi:fracsurfgrowth} %
and propagation of burning fronts%
.\cite{zhang:paperburn}

However if the pinning is strong enough then any elastic structure
will break up and the transport will become inhomogeneous with, at
least in some regimes, flow occuring along channels determined by the
spatial disorder of the random medium. This type of plastic or channel
flow may be found
for strongly pinned flux lines in (at least) thin film superconductors%
,\cite{jensen:plaslett,jensen:plaslong}
for magnetic bubble motion in the presence of impurities%
,\cite{seshadri:bubrev} %
for invasion of non-wetting fluids into porous media%
\cite{%
lenormand:liqporous,
lenormand:invdim}   
(at large enough values of the contact angle%
),\cite{%
cieplak:dynqstatic,
martys:cpfipm,
martys:scaledis,
koiller:fluidsquare,
kushnick:weteffect} 
for fluid flow down a rough or dirty surface%
,\cite{narayan:nlflow,tomassone:unpub}
for charge transport in small metallic dots%
,\cite{middleton:dotlett} %
for dielectric breakdown%
,\cite{niemeyer:breakdown} %
and for vortex motion in disordered Josephson junction arrays%
.\cite{dominguez:djjacp} 

The driven transport of elastic media has been extensively studied in
recent years.  Much less is known about channel flow, and previous
analytic work%
\cite{narayan:nlflow} %
was restricted to the transport of a continuous fluid and concentrated
primarily on the behavior below threshold.  Here, we consider the
motion of discrete particles pushed by a driving force through a
random medium in the limit that the interactions among the particles
essentially just cause constraints on the local number density.

One of the systems of primary interest is vortex motion in strongly
disordered thin film type-II superconductors.  Promise of experimental
observations of the dynamics of this system was sparked by recent
experiments that imaged the vortex lines directly in
real-time%
.\cite{harad:vortexmicroscopy} %
With very strong pinning the intervortex lattice correlations will be
destroyed and, in a thin film, the vortices will behave roughly as
point objects. The driving force is provided by an applied transport
current and we will particularly be interested in the behavior just
above the critical current for the appearance of steady-state vortex
flow. The spatial patterns of the vortex flow, their reproducibility
in a given sample, and various kinds of transients will be the focus
of our attention.

Although the simple models that we introduce are mostly applicable for
very strong pinning, we will argue that on long length scales, the
breaking of the lattice which occurs at large scales
even for weak pinning%
,\cite{coppersmith:instability} %
may lead to qualitatively similar behavior---in particular the
concentration of the vortex flow near the critical current into
relatively narrow, well separated channels.

\subsection{Outline}
\label{sec:outline}

The remainder of the Introduction introduces a very simple model and
discusses some of the qualitative types of phenomena that it exhibits.
Various variants of the model and its behavior above threshold are
analyzed in a mean field approximation in Sec.~\ref{sec:ssmf} and in
two dimensions in Sec.~\ref{sec:ssexact}. In Sec.~\ref{sec:below} some
aspects of the transient behavior below threshold are discussed.
Transients and history dependence are considered in
Sec.~\ref{sec:transhist}. Applications to experimental systems and
conclusions make up Sec.~\ref{sec:conclusions}. Some of the technical
details of the mean field theory below threshold are relegated to the
Appendix.

\subsection{Models}
\label{sec:models}

We introduce a simple model with several features which can be changed
and simplified to allow more information to be obtained.  Somewhat
different variants are convenient for investigating different aspects
of the problem.

The basic model consists of particles on a lattice which move in
downhill directions from site to site. The randomness of the medium is
represented by a random capacity for each site---the number of
particles it can hold without `overflowing'---and a random local rule
for which sites the particle will move to when a given site overflows.
Most of the discussion will be about two dimensional systems or mean
field approximations for these.  We thus, for simplicity, focus on a
square lattice.  One of the diagonal directions of the lattice is
designated as the `downhill' direction. The applied force acts in this
direction which will be referred to as `vertical'. Directions
perpendicular to the vertical will be called `transverse'.  Particles
are only allowed to move between nearest neighbor sites along the two
possible downhill directions. Therefore each lattice site has two
inlets and two outlets as shown in Fig.~\ref{fig:paths}.

If the number of particles on a site exceeds its capacity, then the
excess particles will move out. If there is just one excess particle
then it will overflow through the primary outlet path of the site
which is fixed independently for each site to be to the left or the
right. This is the essential feature of the local randomness in the
medium.  The case in which more than one excess particle is present
will be discussed shortly.

The simplest dynamics are synchronous. Every site on the lattice is
simultaneously updated. At any site where the number of particles
present exceeds the local capacity, those excess particles are moved
along one of the two outlet paths into one of the sites in the next
row of the lattice downhill.  This synchronous dynamics has the
special feature that if all the sites are either full or overflowing,
the below capacity particles are inert while the excess particles in
one row all move down to the next row together and do not interact or
mix with the excess particles in other rows.  It would be more
realistic to allow somewhat asynchronous dynamics.  This would result
in variations in the vertical speed of the excess particles and hence
mixing of the rows. Some features of this will be discussed in
Sec.~\ref{sec:exten}, but for the most part we will study the
synchronous dynamics because of its simplicity.

When particles arrive at a site two things can happen. First, if the
site is not full, some or all of the particles can be trapped at that
site. Particles that are not trapped will, at the next time step,
leave the site through one of the two outlets. If more than one
particle leaves, they can either all leave through the primary outlet
or the particles can be divided among the two outlets, a {\em split}.
The way that this choice is decided and the manner in which the
particles split is governed by the local splitting rule and the
distribution of its parameters.

A simple rule would be to choose the {\em barrier height\/} of each of
the two outlets from some distribution. If the number of particles is
less than the lower outlet barrier, then no particles leave.  If the
number of particles at the site exceeds only one of the outlet heights
then all of the particles above the lower outlet's height leave by
that outlet, i.e., the lower outlet barrier determines the primary
outlet.  Conversely, if the number of particles exceeds the height of
both outlets then the number above the higher outlet could, for
example, be evenly divided between both outlets, with an extra one
going out the higher outlet if the number that exceed the higher
barrier is odd.

Although this splitting rule is suitable for use in simulations, for
analytic calculations it is useful to simplify further. This will be
done in different ways for various calculations. The important aspects
to preserve are that the rule is influenced by the quenched randomness
of the outlet connections---particularly by which is the lower
outlet---and that a site with more excess particles is more likely to
split them among the two outlets

The usual initial condition will be to place particles randomly at
each site. Only the number of particles at each site relative to its
capacity can play a role. We can thus choose this independently for
each site, from a distribution that, in general, will have weight for
both positive (excess) and negative (unfilled) values.  If we are
studying the steady-state behavior above threshold, then it will
sometimes be convenient to only allow sites to start at capacity or
above. This removes the site-filling transients and simplifies the
behavior.

The basic two dimensional model is defined concretely as follows: The
horizontal rows of a square lattice rotated at 45$^\circ$ are denoted
by~$y=1,2,3\ldots$, numbering from the top down, and the sites on
row~$y$ by~$(x,y)$, with~$x+y$ an even integer (see
Fig.~\ref{fig:paths}).  The (integer) number of particles~$n(x,y,t)$
on site~$(x,y)$ at time~$t$ is measured relative to the lower outlet
barrier so that~$n<0$ implies that the site is unsaturated. The height
of the higher outlet barrier relative to the lower barrier is denoted
by an integer~$B(x,y,t)$.

If~$0<n(x,y,t)\leqslant B(x,y,t)$ then at time~$t+1$, $n(x,y,t)$
particles are moved to the site in the next lower row to which the
randomly chosen (but fixed) lower (primary) outlet connects, i.e.,
to one of the sites~$(x\pm1,y+1)$.

If~$n(x,y,t)>B(x,y,t)$ then some fraction of the excess particles
($n_{L}$) are moved through the primary, lower outlet and the
rest,~$n-n_{L}$, through the secondary outlet with higher barrier. The
rule mentioned above corresponds to
\begin{equation}
n_{L}=\left\{
\begin{array}{cl}
  n
  &\quad\text{if $n\leqslant B$}\\
  B+\frac{n-B}{2}
  &\quad\text{if $n-B$ is positive and even}\\
  B+\frac{n-B-1}{2}
  &\quad\text{if $n-B$ is positive and odd.}
\end{array}\right.
\end{equation}

It is best to think of an integer outlet barrier as being the least
integer greater than a continuous valued barrier height. Increasing
the drive~$F$ then corresponds to lowering all of the randomly
distributed barrier heights uniformly, resulting in some fraction of
the integer barrier heights decreasing by one. This is equivalent to
adding a particle to some fraction of the sites, but not quite
equivalent to adding particles randomly, since decreasing all barriers
together should mean that a new particle will not appear a second time
on a site until a new one has appeared on all other sites.
Nevertheless for most purposes the differences will be unimportant

A particularly simple version of the model---essentially the simplest
possible---is the {\em one-deep model}. Here we chose all~$B(x,y)$ to
be equal to one, so that the randomness is only in the choice of which
of the two outlets of each site is the lower, and restrict
the~$n(x,y,t)$ to be no more than two. Then no more than one particle
will be moved through any outlet at a given time step, so
the~$n\leqslant2$ condition will be preserved. Increasing the drive is
equivalent to increasing the number of initial particles. We discuss
some other variants in later sections.

For many purposes we will be interested in large systems so that
steady states can be established very far from the boundaries, but it
is also important to consider the type of boundary conditions
appropriate in a finite system. In particular, unless extra particles
are added at the top of the system all currents will be transient.
Above threshold it is therefore necessary to add particles to the top
row in some way. One possibility is to drive the system at constant
current.  An alternative is to allow all particles moving out from the
bottom row to leave the system and to feed particles into the top row
at the same rate. In either case, the particles added in the top row
could be placed in a regular pattern, a fixed random pattern or in
time-varying random positions. Any of these choices might be
experimentally relevant.

For computer simulations, periodic boundary conditions in the
transverse direction are the obvious choice, so that any particles
leaving the left hand side of the system appear immediately at the
right hand side. It can also be useful to enforce periodic boundary
conditions in the vertical direction giving the system a toroidal
topology, with any particle leaving the bottom row being placed in the
corresponding column in the top row.

Another choice would be to employ open boundary conditions at top and
bottom (no particles enter the top, particles can leave from the
bottom) and to add particles to all sites in the system
stochastically.  This provides a very simple model for rain falling on
a dirty windshield%
.\cite{tomassone:unpub} %
As the system becomes longer in the vertical direction the rate of
rainfall must be decreased concomitantly in order for a well-behaved
large system limit to be reached, we will not study this here.

\subsection{Qualitative Phenomena}
\label{sec:qualphen}

The behavior of even our very simple model is very rich, exhibiting
two distinct `equilibrium' phases, novel critical behavior, as well as
various interesting transient and time dependent effects.

Some of the transient and history dependent effects are very subtle.
We will thus first describe the behavior below threshold and above
threshold in situations where these effects do not play a role.

\subsubsection{Below Threshold}
\label{sec:qualbelow}

If the driving force is low, there will be few excess particles and
they will move downhill readily finding unsaturated sites (`traps') to
fall into and stop. As the force is increased further the extra excess
particles will again fall, but now have to go further before finding
unsaturated sites to fill.

In general this process will be collective and rather complicated:
several particles can end up at the same saturated site at the same
time raising the level there above the higher outlet barrier and hence
splitting the excess particles into two paths. Thus even the route
that a given particle will take depends on the other particles.

There is a simple limit, however, where this does not occur. If
initially there are no excess particles and then the force is
increased very slowly---{\em adiabatically\/}---each excess particle
that appears will be able to fall until it finds an unsaturated site
at which to stop {\em before\/} the next particle appears. The {\em
  path\/} of each particle will then entirely be determined by the
unique primary (lowest) outlet path that emerges from its original
site. Only {\em where\/} on that path it comes to rest will be
determined by the positions of other particles.  Furthermore, the
order in which excess particles appear will not affect the state of
the system after they have all come to rest. The behavior in this
limit is very similar to that in the continuum `river' model below
threshold studied by
Narayan and Fisher (NF)%
.\cite{narayan:nlflow}

The saturated sites and primary outlet paths that connect them will
form tree structures, as shown in Fig.~\ref{fig:trees}, with the
primary outlet of the furthest downhill site of the tree going to an
unsaturated terminus site. Any excess particle that appears on the
tree will fall to this terminus site. If this terminus thereby becomes
filled, the tree will connect to another tree with a terminus further
downhill.

Thus as the drive~$F$ increases, some of the trees will become larger
and larger (even though most will remain small), forming fractals with
typical number of sites scaling with their vertical extent~$L$
as~$L^{d_{f}}$ with~$1<d_{f}<2$ in two dimensions. At a critical
drive~$F_{a}$, corresponding to a critical distribution of the number
of excess particles before they move and the barrier heights, the
vertical length of the longest characteristic trees,
\begin{equation}
  \xi_{-}\sim\frac{1}{\left(F_{a}-F\right)^{\nu_{a}}},
\end{equation}
will diverge, concomitantly the characteristic time scale for most of
the excess particles to come to rest will diverge proportional
to~$\xi_{-}$, since it takes a time~$\xi_{-}$ to fall down a tree of
length~$\xi_{-}$.

The statistics of the saturated trees should be in the same
universality class as in the continuum river network case studied by
NF\@. In the Appendix the mean field analysis is
explicitly carried out for our discrete particle model and compared
with the continuum fluid results.

In two dimensions, NF's numerical results and scaling laws yield
\begin{mathletters}
  \label{eq:nfdata}
  \begin{eqnarray}
    \label{eq:nfnum}
    d_{f}  &=& 1.21\pm0.02\\
    \nu_{a}&=& 1.76\pm0.02.
  \end{eqnarray}
\end{mathletters}

Unfortunately in NF's paper, an erroneous argument was given that the
width~$W$ of a tree of length~$L$ would scale as~$L^{1/2}$. This, by
analogy with conventional directed percolation,
may well not be correct%
.\cite{domany:privcomm} %
More generally we expect
\begin{equation}
  W\sim L^{\alpha_{a}}. 
\end{equation}
The numerics suggest that $\alpha_{a}$ is close to~$1/2$. Thus the width
of the largest characteristic clusters diverges as
\begin{equation}
  W\sim(F_{a}-F)^{-\alpha_{a}\nu_{a}}
\end{equation}
near the threshold.  Any error in assuming $\alpha_{a}=1/2$ would
mean that the error bars quoted in~Eq.~(\ref{eq:nfdata}) are too
small. The numerics were done in a way that makes the true error bars
difficult to reanalyze.

\subsubsection{Flowing Phase}
\label{sec:qualflowing}

The adiabatic approximation breaks down at threshold. If more
particles are added, some of them will start colliding with each other
once they have been `collected' by the large trees. This will result
in splits, forming new outlet paths, filling in of extra sites, and,
eventually, in the formation of a network of persistent `channels'
that drain the system.  The behavior of the transients and the
formation of this network we be discussed later, we first consider the
behavior after the transients have decayed. An example persistent
channel network is shown in Fig~\ref{fig:nodes}.

The simplest way to establish the flowing phase is to insert, at each
time step, particles into the top row of the system with a linear
density~$J$, thereby fixing the input current density. We first
consider a time independent source--a fixed set of sites in the top
row each receive the same number of particles at each time step. If
there were no traps (unsaturated sites) the behavior would be a time
independent pattern of channels after the initial input has passed a
given row. With traps, all the traps on this equilibrium network will
eventually be filled, and afterwards the behavior will again be time
independent, depending only on the input sites and current, the outlet
paths, and the splitting rules. The initial conditions will affect
only the sites that are {\em not\/} on this network. For the time
being we ignore these off-network sites. If the current density is
small, the network will be sparse with only occasional splits and
joins, and most of the outlets will carry a number of excess particles
equal to the minimum above which splits occur.  The characteristic
vertical distance between channel splits and joins in the flowing
phase,~$\xi_{+}$, sets the scale above which the current is
approximately uniform; it diverges as
\begin{equation}
  \xi_{+}\sim\frac{1}{J^2}
  \label{eq:xiplusj2}
\end{equation}
for small current density (in two dimensions). This trivial law
apparently bears no relationship to the scaling properties below
threshold. [Note that~$\xi_{+}$ must be defined rather carefully; the
mean distance between splits is $\text{O}(1/J)$, see
Sec.~\ref{sec:exone}.]

For the simple two-dimensional one-deep model, the statistics of the
channels far from the top of the system can be worked out {\em
  exactly\/} and turn out to be surprisingly simple. One interesting
feature is that the network turns out to be statistically up-down
symmetric, in striking contrast to the strong directionality evident
in the saturated trees below threshold.

An important question now arises: does the channel network far
from the top of the system depend on which sites the current enters?
Of course if all the current is input at one site, the effects of this
will persist, we thus restrict consideration to comparing different
inputs with the same current density and no long range variations in
the density (beyond those expected from, e.g., randomly chosen input
sites).

We thus explicitly consider two randomly chosen input configurations
in the {\em same\/} system. We show that the difference between the
two sets of excess particles after time~$t$, i.e., when they are a
distance~$t$ below the top row, decays as~$t^{-1}$ in mean field
theory---computed explicitly in Sec.~\ref{sec:ssmf}---and
as~$t^{-1/4}$ in two dimensions, found in Sec.~\ref{sec:ssexact} by
analogy with known results for reaction diffusion systems.  Thus, far
from the top, the {\em channel network\/} for a particular random
medium {\em is unique}, depending only on the current density.

\subsubsection{Transients and History Dependence}
\label{sec:qualtrans}

Other than the two simple limits discussed above---adiabatic addition
of particles below threshold and current driven behavior above
threshold---most of the behavior of the system will depend on both its
initial conditions {\em and\/} its history. Here we summarize some of
the features.

Below threshold some qualitative features can be readily guessed. If
(as in Sec.~\ref{sec:qualbelow}) the excess particles are added one at
a time, the only sites that can possibly become filled by particles
not added directly are those which have at least one inlet---a primary
inlet---which is the primary outlet of a neighbor in the row above.
But if many excess particles are added at the same time, enough can
arrive at one site to cause overflow to the secondary outlet, thereby
moving onto some site which may have no primary inlet. Thus a sudden
increase in the force will result in the excess particles spreading
out more, filling a different set of sites, and tending to go less far
downhill before stopping, than in the adiabatic case.

The filling of sites with no primary inlets (and other extra sites
that will also be filled as a consequence of splitting) will make the
primary outlet trees have less saturated sites and thus be smaller.
This will have the consequence of {\em raising\/} the threshold force,
which will, in general, depend on how it is reached.

A specific procedure is to start with a given configuration with no
excess particles and {\em suddenly\/} raise the force to~$F$,
producing many excess particles. When these particles stop moving,
there will again be a distribution of saturated trees whose properties
can be studied as a function of~$F$. This defines a new below
threshold critical phenomena problem which contains some of the above
threshold physics---splits.  If~$F$ is greater than some
critical~$F_{s}$ then the motion of the excess particles will persist
indefinitely in an infinite system.

Does this eventual steady-state still have unsaturated sites? The
answer to this is by no means obvious, indeed, it depends on defining
the limit carefully. In Sec.~\ref{sec:transabove} we will argue that
indeed for any~$F>F_{s}$ reached by suddenly introducing the excess
particles at time zero, there will still be unsaturated sites at
infinite times and a unique network of channels with time independent
flow. The mean current density~$J$ will then be a function of~$F$
with, one would guess, critical behavior,
\begin{equation}
  J\sim\left(F-F_{s}\right)^{\beta_{s}}.
  \label{eq:currentJ}
\end{equation}

An alternative way of reaching a steady state current is to
adiabatically increase the force, as discussed earlier below
threshold. Again an equilibrium channel network will be formed which,
we will argue, will be identical to that formed by the sudden network
at the same value of the current density (but not the same~$F$). But a
different set of off-channel sites will now be saturated and the
current will not behave in the same way as function of~$F$, perhaps
vanishing near the adiabatic threshold~$F_{a}$ with a different
exponent~$\beta_{a}$.

As mentioned in Sec.~\ref{sec:qualbelow}, much of the behavior on
adiabatically increasing~$F$ {\em below\/} threshold is closely
related to that for the continuum river model discussed in NF\@. We
thus first focus on the behavior above threshold in the presence of a
fixed current density.

\section{Mean Field theory of the Flowing Phase}
\label{sec:ssmf}

In many equilibrium and nonequilibrium contexts, various kinds of mean
field approximations provide good starting points. In NF it was found
that for continuum models below threshold, mean field theory was very
instructive but rather subtle. Here we treat the flowing phase of the
discrete particle system in a similar mean field manner focusing on
the behavior of the channel network. For simplicity we assume that all
sites are saturated or have excess particles distributed with mean
density~$J$. This should apply to the channels after the trapping
transients have decayed away.

The crucial feature of the mean field approximation that makes it
tractable is to assume that the inputs to any site are {\em
  uncorrelated}. This should be valid, at least formally, if the
inter-row connections can be made over arbitrarily long transverse
distances instead of between nearest neighbors. We expect also that it
will yield correct critical exponents in sufficiently high dimensions.

One row of excess particles will be followed as it moves down the
system at constant velocity (set to unity). Because only one row is
treated, each lattice site is encountered only once. This means that
the fixed, quenched nature of the outlet connections is irrelevant and
excess particles within the row join and split stochastically. For
convenience, the stochastic dynamics of the joining and splitting is
considered in continuous time rather than using discrete time steps.
Because of its similarity to an aggregation model%
\cite{chandra:agg} %
this approximation will be referred to as the Smoluchowski
approximation.

\subsection{Smoluchowski Mean Field}
\label{sec:smolsmol}

Each site can hold an unrestricted number of particles. It will be
convenient to refer to a fixed size of each particle,~$b$, to
expedite later comparison with flow of a continuous fluid by taking
the limit~$b\rightarrow0$.

The state of the row being followed is fully described, because of the
assumption of the independence of the inlets, by the function~$c(nb,t)$,
the concentration of sites that hold~$n$ particles at time~$t$
within the row.

The concentration of sites that have one or more excess particles
(occupied sites) is described by the function
\begin{equation}
  \phi(t)\equiv b\sum_{n=1}^\infty c(nb,t).
\end{equation}
The factor of~$b$ is inserted so that the sum becomes an integral
and~$\phi$ remains finite in the~$b\rightarrow0$~limit.  The current
density is given by
\begin{equation}
  J(t)\equiv b\sum_{n=1}^\infty nb\,c(nb,t)
\end{equation}
and is trivially equal to the mean number of moving particles per
lattice site.

Smoluchowski dynamics are formulated in continuous time. At most times
all the particles move down the primary outlet paths. Two types of
`reaction' are possible, particles from two sites can join together
and move to a single site (a site to which both inlets are primary)
and a single site can split its particles among two adjacent sites,
corresponding to overflow through the secondary outlet.

Joining is described by the reaction
\begin{equation}
[n]+[m]\longrightarrow[n+m].
\label{eq:reaca}
\end{equation}
The notation~$[n]$ refers to a site with~$n$ particles.  This occurs
at a rate~$\frac{1}{2}\kappa b c(nb)c(mb)$ for each ordered pair~$(n,m)$.

Splitting involves reactions of the form
\begin{equation}
[n]\longrightarrow[m]+[n-m],
\label{eq:reacb}
\end{equation}
where~$m$ is an integer and~$0<m<n$. A simple choice is for each of
the possible splitting reactions to occur at a rate~$\frac{1}{2}\mu b
c(nb)$.  This choice has the desired feature that a site with a larger
number of particles is more likely to split (because of the larger
number of accessible values for~$m$). The total splitting rate for
sites with~$n$ particles will be~$\frac{1}{2}\mu(n-1)bc(nb)$.

The presence of empty sites in these reactions is not included
directly, this is reasonable when occupied sites are dilute. For
regimes where almost all sites are occupied it might be more realistic
to consider splitting terms proportional to~$c(n)c(0)$ to reflect the
fact that an empty site is required for the splitting reaction since
$[n]\rightarrow[m]+[n-m]$ is really $[n]+[0]\rightarrow[m]+[n-m]$.  We
will not analyze this complication here.

The constants~$\kappa$ and~$\mu$ control the overall time scale and
the relative ease of joining and splitting.  The factors of~$b$ ensure
that the total rates are finite in the continuous fluid
limit~$b\rightarrow0$. For convenience we set~$b=1$ henceforth
until we need to consider the limit~$b\rightarrow0$
(Sec.~\ref{sec:contcomp}).

The reactions in Eq.~(\ref{eq:reaca}) and\ (\ref{eq:reacb}) combine
to give the evolution equation for~$c(n,t)$,
\begin{eqnarray}
  \label{eq:cxtevol}
\frac{\partial}{\partial t}c(n,t)&=&
\frac{\kappa}{2}\sum_{m=1}^{n-1}c(m)c(n\!-\!m)
-\kappa c(n)\sum_{m=1}^\infty c(m) \nonumber\\
&-&\frac{\mu}{2}(n-1)c(n)
+\mu\!\sum_{m=n+1}^\infty c(m).
\end{eqnarray}

The parameter~$\mu$ sets the basic (inverse) time scale while the
ration~$\mu/\kappa$ sets the overall scale for the concentrations. We
rescale the concentrations,~$c\rightarrow c\mu/\kappa$, and
times,~$t\rightarrow t/\mu$ to get rid of~$\mu$ and~$\kappa$.

In terms of the transformed variable
\begin{equation}
  \label{eq:transdefn}
  \gamma(\lambda,t)\equiv\sum_{n=1}^\infty e^{-\lambda n} c(n,t),
\end{equation}
the evolution equation now becomes
\begin{equation}
\frac{\partial\gamma}{\partial t}
=\frac{1}{2}\gamma^2
-\phi(t)\gamma
+\frac{1}{2}\frac{\partial\gamma}{\partial \lambda}
+\frac{1}{2}\gamma
+\frac{e^{-\lambda}\phi(t)-\gamma}{1-e^{-\lambda}}.
\label{eq:gammapde}
\end{equation}
Given a knowledge of~$\gamma(\lambda,t)$, it is easy to obtain the
concentration of occupied sites,~$\phi(t)=\gamma(0,t)$, and
the current density
\begin{equation}
  J(t)=-\left.\frac{\partial\gamma(\lambda,t)}{\partial
    \lambda}\right|_{\lambda=0},
\end{equation}
which is conserved.

A simpler measure of the evolution of the system is provided by the
concentration of occupied sites~$\phi(t)$. This function is also
needed to construct an explicit differential equation
for~$\gamma(\lambda,t)$.  The evolution of~$\phi(t)$ is given by
\begin{equation}
\frac{d\phi}{dt}=-\frac{1}{2} \phi^2+\frac{1}{2}(J-\phi),
  \label{eq:mode}
\end{equation}
which is the same as the~$\lambda\rightarrow0$ limit of
Eq.~(\ref{eq:gammapde}).

The steady-state value~$\phi_{\text{eq}}$ is given by the
condition~$d\phi/dt=0$ and has positive solution
\begin{equation}
\phi_{\text{eq}}=\frac{1}{2}\left(
\sqrt{1+4J}-1\right).
  \label{eq:moinfty}
\end{equation}
Equation~(\ref{eq:mode}) is a separable first order equation which
has solution
\begin{equation}
  \frac{\phi(t)-\phi_{\text{eq}}}{\phi(0)-\phi_{\text{eq}}}=
  \frac{2\phi_{\text{eq}}+1}{%
    \left(\phi(0)+\phi_{\text{eq}}+1\right)
    e^{(\phi_{\text{eq}}+1/2)t}
    -\phi(0)+\phi_{\text{eq}}}.
  \label{eq:Mot}
\end{equation}
The asymptotic approach of~$\phi(t)$ to~$\phi_{\text{eq}}$
is given by
\begin{equation}
  \frac{\phi(t)-\phi_{\text{eq}}}{\phi(0)-\phi_{\text{eq}}}\approx
  \frac{2\phi_{\text{eq}}+1}{\phi(0)+\phi_{\text{eq}}+1} e^{-t/\tau_{1}},
  \label{largetMo}
\end{equation}
where the time scale is
\begin{equation}
  \label{eq:motimescale}
  \tau_{1}=
  \frac{1}{\phi_{\text{eq}}+1/2}=
  \frac{2}{\sqrt{1+4J}}.
  \label{eq:smoltau1}
\end{equation}
Observe that, perhaps surprisingly, this timescale is finite in the
limit~$J\rightarrow0$.

The steady state solution for~$\gamma$ is found by taking
$\partial\gamma/\partial t=0$ in
Eq.~(\ref{eq:gammapde}). This gives a Ricatti equation
for~$\gamma_{\text{eq}}(\lambda)$,
\begin{displaymath}
-\frac{1}{2}\frac{d\gamma_{\text{eq}}}{d\lambda}=
\frac{1}{2}\gamma_{\text{eq}}^2
-\left(\phi_{\text{eq}}+\frac{1}{2}+\frac{1}{e^{\lambda}-1}
\right)\gamma_{\text{eq}}
+\frac{\phi_{\text{eq}}}{e^{\lambda}-1}
\end{displaymath}
The solution must satisfy the boundary condition
\begin{equation}
\left.\frac{d\gamma_{\text{eq}}}{d\lambda}\right|_{\lambda=0}=-J
\end{equation}
and respect the symmetry
\begin{equation}
  \gamma\left(\lambda+2\pi i\right)=\gamma(\lambda)
  \label{eq:discsym}
\end{equation}
required by the restriction that~$c(n)$ be non-zero only for~$n\in
\{1,2,3\ldots\}$.

The only suitable solution is thus
\begin{equation}
   \gamma_{\text{eq}}(\lambda)=\frac{\phi_{\text{eq}}}{%
     \left(\phi_{\text{eq}}+1\right)e^{\lambda}- \phi_{\text{eq}}},
  \label{eq:ginf1}
\end{equation}
whose transform yields the steady state solution
\begin{equation}
  \label{eq:cinf}
  c_{\text{eq}}(n)
  =\left(\frac{\phi_{\text{eq}}}{\phi_{\text{eq}}+1}\right)^{n}
    =\left(\frac{\sqrt{1+4J}-1} {\sqrt{1+4J}+1}\right)^{n}.
\end{equation}

This can be written
\begin{equation}
  c_{\text{eq}}(n)= e^{-n/\Upsilon},
  \label{eq:cinfexp}
\end{equation}
with characteristic scale for decay of the number of particles at an
occupied site
\begin{equation}
  \Upsilon=\frac{1}{\ln \left(1+1/\phi_{\text{eq}}\right)}.
  \label{eq:charnum}
\end{equation}

Although the general dynamics of~$\gamma(\lambda,t)$ do not satisfy
detailed balance, detailed balance is recovered in the steady state
distribution.  The rate at which particles move from sites with~$n$
particles to sites with~$m>n$ particles is
\begin{equation}
  \Gamma_{n\rightarrow m}(t)=\frac{n}{2} c(n,t)c(m-n,t).
\end{equation}
The reverse rate of transfer is
\begin{equation}
  \Gamma_{m\rightarrow n}(t)=\frac{n}{2} c(m,t).
\end{equation}
These two rates are equal with the equilibrium form
of~$c_{\text{eq}}(n)$ shown in Eq.~(\ref{eq:cinf}).

This suggests that there is substantial simplicity in the steady
state.  We will see later that, in fact, in a truncated model, all
allowed configurations are equally likely.

The discrete nature of the particles is most important in the limit of
a small current,~$J\rightarrow0$.  In this limit the steady-state
current distribution is dominated by sites with only a single
particle. The concentrations for sites with large numbers of particles
become exponentially small, dropping by a factor of approximately~$J$
for each additional particle
\begin{equation}
  \frac{c_{\text{eq}}(n+1)}{c_{\text{eq}}(n)}=J+\text{O}(J^2)
\label{eq:addpart}
\end{equation}
and correspondingly
\begin{equation}
  \Upsilon\sim\frac{-1}{\ln J}.
\end{equation}

As~$J\rightarrow0$, the fraction of particles on sites with only a
single particle approaches unity like
\begin{equation}
  \frac{c_{\text{eq}}(1)}{\sum_{n}^\infty n c(n)}=
  1-2J+\text{O}(J^2).
\end{equation}
It follows that the number of occupied states must be be proportional
to the current,
\begin{equation}
  \phi_{\text{eq}}\sim J+\text{O}(J^2).
\end{equation}

The correlation length~$\xi_{+}$ can be defined as the vertical
distance over which the current is correlated. In mean field, this is
the same as the typical downhill distance between splits in the
channel network, so the rate of splitting is inversely proportional
to~$\xi_{+}$, just as in NF.

The rate for splitting at any site is given by
\begin{equation}
  \frac{1}{2}\sum_{n=1}^\infty (n-1)c(n)=\frac{1}{2}(J-\phi_{\text{eq}})
  =\frac{1}{2}\phi_{\text{eq}}^2.
\end{equation}
The rate for splitting among occupied sites is
therefore~$\frac{1}{2}\phi_{\text{eq}}$ and thus, with the particle
velocity set to unity, the typical distance a particle travels before
reaching a splitting site of the network is
\begin{equation}
  \xi_{+}=\frac{2}{\phi_{\text{eq}}}\sim\frac{1}{J}
  \label{eq:xiprimemoi}
\end{equation}
in the small current limit.

The relaxation time for approach to the steady state is given by
\begin{equation}
  \tau_{1}=2-4J+\text{O}(J^2).
  \label{eq:smoltau2}
\end{equation}
This approaches a finite limit as~$J\rightarrow0$. The limiting
time-scale is set by the splitting rate [$1/\mu$ in
Eq.~(\ref{eq:cxtevol})].  No joining is necessary for equilibration in
this mean field approximation. The finite relaxation time is obtained
because, although~$dc(n,t)/dt$ vanishes as~$J^n$, the equilibrium
concentration~$c_{\text{eq}}(n)$ is also of order~$J^n$

The finite relaxation time is only for the equilibration of the
distribution of excess particle numbers. The time for spatial
correlations of the equilibrium network to develop {\em does\/}
diverge as~$J\rightarrow0$, but this is not seen by studying only a
single row. We will return to this in Sec.~\ref{sec:smolconv}.

\subsection{Comparison with Continuous Fluid}
\label{sec:contcomp}

A mean field model with a continuous fluid, like those studied by NF,
can be recovered by reinstating the particle size~$b$ and then
taking the limit~$b\rightarrow0$. The results obtained for a small
current are then {\em different\/} from the discrete case, that is the
limits~$J\rightarrow0$ and~$b\rightarrow0$ do not commute.

In the continuum limit the concentration~$c(x)$ is a function of a
continuous variable and the sum~$\phi$ is replaced by the integral,
\begin{equation}
 \phi(t)=\int_{0}^\infty\!dx\,c(x,t)
 =\lim_{b\rightarrow0}\sum_{n=1}^\infty b\,c(nb,t)
\end{equation}
and the current is defined by
\begin{equation}
 J(t)=\int_{0}^\infty\!dx\,x\,c(x,t)
 =\lim_{b\rightarrow0}\sum_{n=1}^\infty nb^2\,c(nb,t).
\end{equation}

The steady state value of~$\phi$ is
\begin{equation}
  \phi_{\text{eq}}
  =\lim_{b\rightarrow0} \frac{b}{2}
  \left(\sqrt{1+\frac{4 J}{b^2}}-1\right)
  =\sqrt{J}
\end{equation}
This value of~$\phi_{\text{eq}}$ shows that the fraction of sites that
are in channels vanishes much more quickly for small current in the
continuous case ($\phi\sim J^{1/2}$) than in the discrete case
($\phi\sim J$). This is because in the discrete case the mean number
of particles at an occupied site ($J/\phi$) cannot be less than one,
whereas in the continuous case the mean amount of fluid at each site
approaches zero as~$J\rightarrow0$.

The steady-state site concentration distribution can be
obtained from the transform
\begin{equation}
  \gamma_{\text{eq}}(\lambda)=\lim_{b\rightarrow0}
  \frac{b\phi_{\text{eq}}}{\left(\phi_{\text{eq}}
    + b\right)e^{\lambda b} -\phi_{\text{eq}}}
  =\frac{1}{\lambda+1/\phi_{\text{eq}}},
\end{equation}
yielding
\begin{equation}
  c_{\text{eq}}(x)= \exp\left(-x/\sqrt{J}\right),
\end{equation}
so that the typical channel site has~$\sim\sqrt{J}$ excess
fluid.

The vertical correlation length can also be found in the continuum limit:
\begin{equation}
  \xi_{+}=\frac{2}{\phi_{\text{eq}}}
  =\frac{2}{\sqrt{J}};
\end{equation}
it diverges more slowly than Eq.~(\ref{eq:xiprimemoi}),
where~$\xi_{+}\sim1/J$. Finally the time scale for equilibration of
the single site properties, such as~$\phi$, in a continuum fluid is
\begin{equation}
  \tau_{1}=\frac{1}{\phi_{\text{eq}}},
\end{equation}
which diverges as~$J\rightarrow0$ in contrast to the discrete case.

In the continuous fluid,~$\xi_{+}$ and $\tau_{1}$ both diverge
as~$J^{-1/2}$. This divergence should be expected: as~$J\rightarrow0$
the rivers get shallower and split less often, increasing the size of
the flow pattern ($\xi_{+}$) and requiring longer to equilibrate
($\tau_{1}$).  In contrast, in the discrete case~$\xi_{+}$ diverges
simply (as $J^{-1}$) as the channel sites get further apart, but the
time scale,~$\tau_{1}$, found above does not diverge. If we consider,
for example, an initial condition where there are single isolated
particles on small fraction~$J$ of the sites, then the steady state
number of sites with each occupation has almost been reached,
so~$\tau_{1}$ is small and finite, and the vertical length scale for
the flow pattern is~$J^{-1}$, the joining rate.

Note that the behavior of the continuous fluid in the limit of small
current depends (in contrast to the discrete case) on the splitting
rule. The splitting rule used here and in NF are thus appropriate for a
distribution of the outlet barrier differences~$B$ which is constant
near~$B=0$. A different form of the distribution for small~$B$ will
result in different exponents characterizing the small current
behavior since the splitting rate of shallow rivers will be a
different function of their depth. This is somewhat analogous to the
dependence of certain exponents in conventional continuum percolation
on local properties of the medium%
.\cite{halp:contperc+feng:contperc}

\subsection{Restricted Depth Models}
\label{sec:truncmf}

In the Smoluchowski mean field calculation we found that the
steady-state concentration of sites containing~$n$ particles scales as
[Eq.~(\ref{eq:addpart})]
\begin{equation}
c(n)\sim J^n,
\end{equation}
when the mean current of particles~$J$ is small. This suggests that,
at small~$J$, a reasonable approximation will be to limit the
possible number of particles at each site to
\begin{equation}
  0\leqslant n\leqslant N,
\end{equation}
for some constant value N.

This restriction  makes theoretical treatment of the model much
simpler.

The `depth' restriction can conveniently be enforced for even values of~$N$
by allowing no more than~$N/2$ particles to travel down any outlet
in one time-step. We will consider two versions, a `one-deep' model,
which allows only one particle to travel through any outlet,
and a `two-deep' model, which allows two.

For the {\em one-deep model}, mentioned in the Introduction, each site
can hold at most two particles (i.e.,~$N=2$). This is preserved by using
synchronous dynamics and allowing no more than one particle to pass
along any outlet. If a site holds one particle, then the particle
moves onto the next row through the primary outlet. If
the site has two particles, one particle moves through each of the two
outlets.

The rules are similar for the {\em two-deep model}. A site can have up
to four particles ($N=4$) and each outlet can pass a maximum of two
particles.  A site with one particle passes that particle to the
primary outlet. A site with four particles passes two particles to
each outlet. A site with three particles passes two to the primary
outlet and one to the secondary outlet. A site with two particles can
either pass both particles to the primary outlet or it can pass one
particle to each of the two outlets. This choice is fixed
independently for each site with probability~$r$ of the site sending
both particles through the same outlet. The choice~$r=1/3$ makes the
mean-field calculation simpler and will be treated separately.

Both of these models have a particle-hole symmetry. A near-maximal
current behaves the same way as a very small current. The truncation
that creates these models is of course only reasonable in the latter
case.

As for the Smoluchowski mean field, we choose a particular row of
particles and follow that row as it moves down through the system.
The probability that a site in that row holds~$n$ excess particles at
time~$t$, is given by~$p(n,t)$.  With synchronous, discrete time
dynamics the probabilities at time~$t+1$ depend only on those at
time~$t$, so we will construct a set of recursion relations
giving~$p(n,t+1)$ in terms of~$p(n,t)$. For brevity we will write
these as~$p'_{n}$ and~$p_{n}$ respectively.

To calculate the values~$p'_{n}$ for a given site, X, we need to look
at the two sites (say V~and~W) which provide inputs to X. The inlets
VX~and~WX are each equally likely to be primary or secondary outlets,
so each combination of input types occurs with probability one
quarter.  The simplest way to calculate~$p'_{n}$ for X is to consider
every possible combination of inlet types and of occupations for
V~and~W.  For each combination the new occupation of X is given by the
splitting rules discussed above.

\subsubsection{One-deep}

For the one-deep case, we obtain
\begin{mathletters}
\begin{eqnarray}
  p'_{0} &=& p_{0}^2+p_{0} p_{1}+\frac{1}{4}p_{1}^2 \\
  p'_{1} &=& p_{0} p_{1} +p_{1}p_{2} +2p_{0}p_{2}+\frac{1}{2}p_{1}^2\\
  p'_{2} &=& p_{2}^2+p_{1}p_{2}+\frac{1}{4}p_{1}^2.
\end{eqnarray}
\end{mathletters}

The analysis can be simplified by using the probabilities~$q_{n}$
of~$n$~($\!=\!0$ or 1)~particles passing through an arbitrary primary
or secondary inlet to site X in one time step:
\begin{mathletters}\begin{eqnarray}
p_{0}&=&q_{0}^2\\
p_{1}&=&2q_{0}q_{1}\\
p_{2}&=&q_{1}^2,%
\end{eqnarray}\end{mathletters}%
and, for an outlet from site X, which can be either the primary or
secondary outlet (each with probability~$1/2$), we obtain the simple
recursion relations
\begin{mathletters}\begin{eqnarray}
  q'_{0}&=&q_{0}^2+q_{0}q_{1}\\
  q'_{1}&=&q_{1}^2+q_{0}q_{1},%
\end{eqnarray}\end{mathletters}%
where~$q'_{n}$ refer to the outlets of a site and~$q_{n}$ refer to its
inlets. The mean field approximation makes the two inlets independent.
The two outlets from a site are correlated but only one outlet from
each site need be treated here as the two outlets will go to different
(`far away') sites.

For the one-deep case there are two variables and two conserved
quantities: the total probability,~$q_{0}+q_{1}=1$, and
the mean current per site,~$J=2q_{1}$. This means that there
are no dynamics in mean field and, trivially,
\begin{mathletters}
\begin{eqnarray}
  q_{0}&=&\frac{1}{2}(2-J) \\
  q_{1}&=&\frac{1}{2}J.
\end{eqnarray}
\end{mathletters}
\subsubsection{Two-deep}

The recursion relation for the two-deep model can be derived in the
same way as the one-deep model, again working in outlet variables,
\begin{mathletters}
\begin{eqnarray}
q'_{0}&=&q_{0}^2+q_{0} q_{1} +\frac{1}{2}q_{1}^2r+q_{0}q_{2}r\\
q'_{1}&=&q_{0}q_{1}+q_{1}^2(1-r)+2q_{0}q_{2}(1-r)+q_{1}q_{2}\\
q'_{2}&=&q_{2}^2+q_{1} q_{2} +\frac{1}{2}q_{1}^2r+q_{0}q_{2}r.
\end{eqnarray}
\end{mathletters}
These can be reduced to the single relation
\begin{equation}
  q'_{1}=\frac{1}{8}J(4-J)(1-r)+\frac{1}{2}q_{1}^2(1-3r)+q_{1}r,
  \label{eq:01234rec}
\end{equation}
by applying the two conservation laws,
\begin{equation}
q_{0}+q_{1}+q_{2}=1
\end{equation}
and
\begin{equation}
  J=2q_{1}+4q_{2}.
\end{equation}

The two-deep model has one free parameter, the fraction ($r$) of sites
from which, when the site is doubly occupied, both particles will
take the same outlet.

The cases~$r=0$ and~$r=1$ are special. For~$r=0$ the depth-two outlets
eventually disappear and the equilibrium is the same as the one-deep
model. For~$r=1$ all sites eventually have an an even number of
particles and the model becomes a doubled version of the one-deep model.
The case~$r=1/3$ removes the quadratic term from the~$q_{1}$-recursion
relation, simplifying the analysis.

For general values of~$r$, the recursion relation
[Eq.~(\ref{eq:01234rec})] is quadratic
and converges to the fixed point
\begin{equation}
q_{1}^{\text{eq}}=\frac{%
  1-r-\sqrt{%
    (1-r)\left[%
    (1-r)+J(1-J/4)(3r-1)
  \right]}}{%
1-3r}.
\end{equation}

This steady-state distribution has the small current limit
\begin{mathletters}
\begin{eqnarray}
  q_{0}^{\text{eq}}&=&
  1-\frac{1}{2}J+\frac{1}{8}\frac{r}{1-r}J^2+\text{O}(J^3)\\
  q_{1}^{\text{eq}}&=&
  \frac{1}{2}J+\frac{1}{4}\frac{r}{1-r}J^2+\text{O}(J^3)\\
  q_{2}^{\text{eq}}&=&
  \frac{1}{8}  \frac{r}{1-r}J^2+\text{O}(J^3).
\end{eqnarray}
\end{mathletters}

Convergence towards this equilibrium is exponential with time-constant
\begin{equation}
\tau_{1}=\frac{-1}{%
  \ln \left(1-\sqrt{(1-r)\left[(1-r)+J(1-J/4)(3r-1)\right]}\right) }
\label{eq:2dptau}
\end{equation}
which does not diverge in the physical regime ($0\leqslant
r\leqslant1$ and~$0\leqslant J\leqslant4$) except at~$r=1$ and
at~$r=0$,~$J=2$.

For the case~$r=1/3$ the recursion relation has the simple linear form
\begin{equation}
  q'_{1}-q_{1}=\frac{1}{12}J(4-J)-\frac{2}{3}q,
\end{equation}
with a single fixed point at
\begin{equation}
  \left(
    \begin{array}{c}
      q_{0}^{\text{eq}} \\
      q_{1}^{\text{eq}} \\
      q_{2}^{\text{eq}}
    \end{array}
  \right)
  =\frac{1}{16}
  \left(
    \begin{array}{c}
      (4-J)^2 \\ 2J(4-J) \\ J^2
    \end{array}
  \right)
\end{equation}
and convergence is exactly exponential,
\begin{equation}
  q'_{1}-q_{1}^{\text{eq}}=\frac{1}{3}\left(q_{1}-q_{1}^{\text{eq}}\right),
\end{equation}
with time scale~$1/\ln 3$. The independence of~$J$ is special to
this value ($r=1/3$) but the finite convergence rate is more
generic.

\subsubsection{Comparison with Smoluchowski}

We see that the finite time scale for convergence of the single site
distribution exists both for these truncated models and the
Smoluchowski model in the mean field limit.  Other aspects of the
truncated models can be compared with the Smoluchowski mean field.
They should be similar for small currents.

The basic features are easily checked for small~$J$.
The~$c_{\text{eq}}(n)\sim J^n$ property that motivated the truncation
is preserved for both one-deep and two-deep models. In the truncated
models this corresponds to the site variables relation
$p_{n}^{\text{eq}}\sim J^n$ which follows from the fact that the
outlet variables obey~$q_{n}^{\text{eq}}\sim J^{n}$.  Concomitantly,
as~$J\rightarrow0$ the fraction of sites that have exactly one
particle approaches unity linearly with~$J$, but with a model
dependent coefficient.

The correlation length is given by the scale of splitting in the river
network and is inversely proportional to the rate of splitting in
the steady state. For the one-deep model all the sites that have two
particles split so that
\begin{equation}
  \frac{1}{\xi_{+}}\propto\frac{p_{2}}{1-p_{0}}=\frac{q_{1}^2}{1-q_{0}^2}
  \sim J
\end{equation}
as~$J\rightarrow0$ just as in the Smoluchowski case
[Eq.~(\ref{eq:xiprimemoi})]. In the two-deep case the splitting rate
is proportional to~$(1-r)p_{2}+p_{3}+p_{4}$ and is
also~$\text{O}(J)$ for small~$J$.

\subsection{Convergence to Channel Network}
\label{sec:smolconv}

In this subsection we study the development, in mean field theory, of
the channel network. To do this we consider two rows of particles
which start with different initial configurations but move in the {\em
  same\/} realization of the random medium, starting at the same row.
This allows us to address some interesting questions. Does the flow
from different initial conditions approach a fixed network?  If the
system is driven by a current at a top boundary how far down does the
influence of the distribution of the driving current extend?

We will examine the one-deep and two-deep models first.

\subsubsection{One-deep}

In the mean field approximation the state of the two copies of the
system is represented by the joint probability~$p_{mn}$ that a site in
the first copy holds~$m$ particles and a site in the second copy
holds~$n$ particles. For the one-deep system this requires a
three-by-three matrix of probabilities
\begin{equation}
  \left(
    \begin{array}{ccc}
      p_{00} & p_{01} & p_{02} \\
      p_{10} & p_{11} & p_{12} \\
      p_{20} & p_{21} & p_{22}
    \end{array}
  \right).
\end{equation}

Again, it is simpler to use the outlet representation
\begin{equation}
  \left(
    \begin{array}{cc}
      q_{00} & q_{01}\\
      q_{10} & q_{11}
    \end{array}
  \right),
\end{equation}
whose state matrix has four parameters, but only three are independent
because the total probability must be unity. The current density~$J$
which is the same for both copies fixes two combinations of
the~$q_{mn}$ and thus given one of the entries in the state matrix
(say $q_{00}$) we know all of the others.

The recursion relations for this system can be written down in just
the same way as for the single system. Now the iteration is over the
number of particles entering along the inlets in each of the two
systems and over whether the outlet (in both systems together) is
the primary or secondary:
\begin{equation}
  \left(
    \begin{array}{cc}
      q'_{00} & q'_{01}\\
      q'_{10} & q'_{11}
    \end{array}
  \right)
  =
  \left(
    \begin{array}{ccc}
q_{00}+q_{01}q_{10} & \quad & q_{01}-q_{01}q_{10} \\
q_{10}-q_{10}q_{01} & \quad & q_{11}+q_{11}q_{00}
    \end{array}
  \right).
\end{equation}
Using the constraints this reduces simply to
\begin{equation}
  q'_{00}-q_{00}=\left(1-\frac{1}{2}J-q_{00}\right)^2,
  \label{eq:012rec2}
\end{equation}
which yields the simple diagonal steady state matrix
\begin{equation}
  \left(
    \begin{array}{cc}
      q_{00}^{\text{eq}} & q_{01}^{\text{eq}}\\
      q_{10}^{\text{eq}} & q_{11}^{\text{eq}}
    \end{array}
  \right)
  =
  \left(
    \begin{array}{cc}
      1-J/2 & 0 \\
      0 & J/2
    \end{array}
  \right),
\end{equation}
corresponding to each of the two systems being locked in the same
state. Thus, indeed we see that the two initial conditions converge to
the {\em same\/} steady state channel network.

Because the recursion relation [Eq.~(\ref{eq:012rec2})] is a perfect
square, the approach to the steady state is only quadratic. If
$q_{00}=q_{00}^{\text{eq}}-\delta$ then the change~$q'_{00}$-$q_{00}$
is only~$\delta^2$ which means that the difference between~$q_{00}$
and~$q_{00}^{\text{eq}}$ decreases like~$t^{-1}$. This is a much
slower equilibration than for the single site distribution in mean
field where the approach was exponential

\subsubsection{Two-deep Model}

To describe two copies in the two-deep model using the outlet
representation requires a three-by-three matrix
\begin{equation}
  \left(
    \begin{array}{ccc}
      q_{00} & q_{01} & q_{02} \\
      q_{10} & q_{11} & q_{12} \\
      q_{20} & q_{21} & q_{22}
    \end{array}
  \right).
\end{equation}
The recursion relations for the matrix can again be constructed by
enumerating all possible inputs to both systems. The fate of doubly
occupied sites is, of course, chosen once for each lattice site, with
the probability parameterized by~$r$, and thus has the same result in
each of the two copies.

Because, in mean field, each single system equilibrates exponentially,
it is simplest to consider the subspace where each single system is
equilibrated (this happened trivially for the one-deep model by
specifying~$J\/$). This requirement fixes the sum of each row and column
of the state matrix so we have only four remaining degrees of freedom,
say~$(q_{00}, q_{01}, q_{10}, q_{11})$.  The recursion relation is then
between two 4-vectors.
\begin{equation}
  \left(\begin{array}{c}
    q_{00} \\ q_{01} \\ q_{10} \\ q_{11}
  \end{array}\right)
  \longrightarrow
  \left(\begin{array}{c}
  q'_{00} \\ q'_{01} \\ q'_{10} \\ q'_{11}
  \end{array}\right).
\end{equation}
The new values of each coefficient are quadratic functions of the four
previous coefficients and quartic functions of~$J$.

The state where both copies are in the same state is a fixed point of
this relation. To show that this is a stable fixed point requires more
work.  We first consider linear stability analysis with a
perturbation around this fixed point
\begin{equation}
  \left(\begin{array}{c}
  q_{00} \\ q_{01} \\ q_{10} \\ q_{11}
  \end{array}\right)
  =
  \left(\begin{array}{c}
  q^{\text{eq}}_{0} \\ 0 \\ 0 \\ q^{\text{eq}}_{1}
  \end{array}\right)
  +
  \left(\begin{array}{c}
  x_{1} \\ x_{2} \\ x_{3} \\ x_{4}
  \end{array}\right).
\end{equation}
Because the matrix values are probabilities this means, for this fixed
point, only perturbations with~$x_{2}\geqslant0$ and~$x_{3}\geqslant0$
are physical.

We focus on the simple~$r=1/3$ case.  To leading order in the~$x_{i}$
variables, the recursion relation
for~$\vec{x}=(x_{1},x_{2},x_{3},x_{4})$ is
\begin{equation}
  \vec{x}'=M\vec{x}
\end{equation}
where~$24M$ is the matrix
\begin{displaymath}
  \left(\begin{array}{cccc}
  8(5-J) & 4(4-J) & 4(4-J) & 4(4-J) \\
  (J-4)(J+8) & J^2-8 & 4(J-4) & 4(J-4) \\
  (J-4)(J+8) & -4(4-J) & J^2-8 & 4(J-4) \\
  32+8J-2J^2 & 16+4J-J^2 & 16+4J-J^2 & 1
  \end{array}\right)
\end{displaymath}

The matrix~$M$ is not symmetric. It has the eigenvalues~$1$, $1/3$ and
$(J^2-4J+8)/24$, the last being doubly degenerate.  The four right
eigenvectors are not orthogonal but they span the space and any
perturbation can be written as a linear combination of them.  We
write~$\vec{x}$ as a component parallel to the first (right-)
eigenvector and a component perpendicular to its corresponding
left-eigenvector (i.e., as a linear combination of the other three
right-eigenvectors)
\begin{equation}
  \vec{x}=a_{1}\vec{e}_{1}+a_{\perp}\vec{e}_{\perp}
\end{equation}

Except for the first eigenvalue, the others are never greater
than~$1/3$. This means, that if this perturbation is subjected to the
iteration then the perpendicular component will go to zero
exponentially on a time scale of~$1/\ln 3$ or less.  The long time
behavior is therefore controlled by the eigenvector corresponding to
the unit eigenvalue. To linear order the unit eigenvalue indicates
that the perturbation is marginal. Stability can be determined by
projecting the full recursion relation onto the direction of the first
(right-) eigenvector using the corresponding left-eigenvector.  This
yields
\begin{equation}
  a'_{1}=a_{1}-a_{1}^2.
\label{eq:recurA}
\end{equation}
This implies that~$a_{1}$ decays to zero as~$t^{-1}$ whenever~$a_{1}$
is positive.  Formally the fixed point is unstable to
negative~$a_{1}$, however this is unphysical because the eigenvector,
\begin{equation}
  \label{eeeone}
  \vec{e}_{1}=
      \left(\begin{array}{c}
    -1+J/4 \\ 1-J/4 \\ 1-J/4 \\ -1
    \end{array}\right),
\end{equation}
then produces negative entries in the probability matrix. Thus the
fixed point with the two copies locked together is marginally {\em
  stable}.

For general values of~$r$ the behavior is very similar to that for the
case~$r=1/3$ with the same recursion relation for~$a_{1}$.  The system
will again approach the state where both copies are in the same state,
but will do so only as~$t^{-1}$.

\subsubsection{General Mean field}

The mean field treatment of the two simple models showed that the flow
pattern will approach a fixed network but only algebraically with the
fraction of sites for which the initial conditions alter the network
decaying at long times as~$t^{-1}$.  A general argument, based on
analogy with a reaction diffusion system, shows that in mean field
theory this result is general.

The decay can be understood by keeping track of inlets in which the
occupation in the two copies differ. For the one-deep model we label
each inlet with an extra particle in the first copy `A' and each inlet
with higher occupation in the second copy `B'. Inlets which have the
same occupation in each copy are not labeled. As the row is followed
down the system, the A's and B's move independently (except that no
more than one particle may occupy a site) until an A meets a B. The
corresponding particles from then on follow the same path and thus the
A and B disappear. This therefore corresponds to an
$A+B\rightarrow\emptyset$ reaction diffusion
system%
.\cite{ovchinikov:reacsem,toussaint:partanti}

In the continuous time approximation (like the Smoluchowski model) the
$A+B\rightarrow\emptyset$ mean field equations are (in dimensionless
form),
\begin{equation}
  \frac{d\rho_{A}}{dt}=\frac{d\rho_{B}}{dt}=-\rho_{A}\rho_{B}.
\end{equation}
with~$\rho_{A,B}$ the concentrations of A's and B's. With the same
current density in each copy we begin with equal concentrations of A
and B, and the solution is
\begin{equation}
  \rho_{A}=\rho_{B}=\frac{1}{t}.
\end{equation}

For the two-deep and more general models, the analysis is rather more
complicated since there are a variety of ways in which the two copies
can differ. These can combine in different ways as well as some
combinations annihilating. In the continuous time
approximation--generally good at long times---there will be a coupled
series of quadratic recursion relations for the
densities,~$\left\{\rho_{i}\right\}$, of sites in which the two copies
differ in various ways. These equations will generally have a solution
with all the~$\rho_{i}$ decaying to zero proportional to~$t^{-1}$. As
for the two-deep analysis at long times we expect the general initial
conditions to decay exponentially to this asymptotic form. Thus the
$t^{-1}$~behavior for convergence of two initially randomly different
copies will be generic in mean field.

\section{Flowing Phase in Two Dimensions}
\label{sec:ssexact}

In this section properties of the current distribution in the flowing
phase are analyzed for the two dimensional systems of primary
interest. Some of the features will be qualitatively similar to those
found in mean field theory in the previous section, but of course
non-trivial correlations will exist that are absent in mean field
theory. We will focus on the one-deep model for which a number of
exact results can be obtained. In Sec.~\ref{sec:exten} the extent to
which the results apply more generally will be discussed.

\subsection{One-Deep Model}
\label{sec:exone}

The one-deep model defined in the Introduction is the simplest model
with the essential features. Each outlet carries either zero or one
particles at each time step so any site can contain at most two
particles above its capacity.

As in the previous section, we will focus on the behavior after all
traps that might be filled have been filled. As discussed in the
Introduction this can most easily be achieved by adding particles at a
fixed rate in the top row. Thus all accessible sites can be considered
to be at or above their capacities.

\subsubsection{Single row statistics}

In mean field theory, the only quantity of interest for the evolution
of a single row moving down the system is the single outlet occupation
distribution which is trivially determined by the current density~$J$.
In two dimensions any translational invariant distribution for the
outlet occupation in a row will again have the same single outlet
distribution. Surprisingly, the steady state row distribution turns
out to have the same simple product form as in mean field so that the
steady state occupation of each outlet in a row is {\em independent}.
This can be seen to be a steady state by considering the two inlets
and two outlets of a given site.

If the two inlets are independently occupied with probability~$J/2$,
then at the next time step, the probability of either both or neither
of the outlets being occupied are trivially~$J^2/4$ and~$(1-J/2)^2$
respectively. Although the behavior if exactly one of the inlets is
occupied is determined by which of the outlets is the primary outlet,
the fact that this is independent from site to site and that each site
is only encountered once as a row of particles moves down the system
immediately means that outlet occupations 10 and 01 are equally likely
with probability~$(1-J/2)J/2$ independently for each pair. Thus, the
outlets are independently occupied if the inlets are, proving that the
independent occupation distribution is a steady state. We expect that
this will be the unique steady state; this should be provable along
the lines of the arguments developed below.

\subsubsection{Uniqueness of the channel network}

Before considering the nontrivial correlations that exist between rows
in the steady state, we first show that, as in mean field, the actual
network of channels---i.e., which outlets are occupied---does not
depend on the initial conditions. We use arguments analogous to those
for the mean field case, considering two copies of the
system---specifically following a given row---with different random
initial conditions. The same strategy can be followed as for the mean
field treatment of Sec.~\ref{sec:smolconv}.

Consider a single inlet on the row to be followed. It contains either
zero or one particles in each copy.  We can represent these possible
states as~$00$, $01$, $10$ and~$11$.  Each outlet, the inlets to the
next row down at the next time step, also contain zero or one
particles in each copy; these are represented similarly.

If one particle enters the first copy of the site through, say, the
left inlet and no particles enter the second copy, then the site will
output one particle in the first copy and no particles in the second
copy. Using our notation for the four sites this can be written
\begin{equation}
  10+00\longrightarrow10+00,
\end{equation}
where the output~$10$ will be, with equal probability, on the left or
right outlet.

If a single particle entered from the same side (e.g., the left) in
both copies then one particle would leave in both copies through the
same outlet. This can be written
\begin{equation}
  11+00\longrightarrow11+00.
\end{equation}

This evolution rule
\begin{equation}
  x+y\longrightarrow x+y
\end{equation}
holds for all but one case. If a single particle enters both copies of
the site, but the particle enters from different inlets in the two
copies, the particle will leave both copies of the site via the same
(primary) outlet. This can be written
\begin{equation}
  10+01\longrightarrow11+00.
\end{equation}

The notation can thus be simplified by considering only the~$10$
and~$01$ outlets, which can conveniently be called A-states and
B-states. Our single non-trivial reaction is then
\begin{equation}
  A+B\longrightarrow\emptyset,
\end{equation}
as in mean field, where~$\emptyset$ represents any combination of~$00$
and~$11$, i.e., sites where both copies have the same occupation.
Apart from this single reaction, the A and B sites perform independent
unbiased random walks along their row with the `excluded-volume'
restriction that no more than one state can be assigned to each
outlet.  This model is therefore an example of the extensively studied
$A+B\rightarrow\emptyset$ reaction diffusion system%
.\cite{ovchinikov:reacsem,toussaint:partanti}

The basic picture which should apply in low dimensions is as follows.
In a time~$t$, the A and B particles will perform random walks in the
transverse direction, annihilating if they encounter each other. Since
in time~$t$, two random walks that start a distance less
than~$\text{O}(t^{1/2})$ apart are very likely to intersect, for long
times the minority species in regions of width less
than~$\text{O}(t^{1/2})$ are likely to be completely annihilated in
time $t$ by the majority species in that region. Thus after time~$t$,
the system will be, roughly, divided into regions of width~$\sim
t^{1/2}$, each of which contains mostly A or mostly B. The number of
A's in an A rich region will be of order the typical excess of A over
B in that region initially, which, for initial conditions without long
range spatial correlations, will be of order~$(t^{1/2}J)^{1/2}$ for
small~$J$. Thus after long time~$t$, the density of A's (or B's)
remaining will be of order
\begin{equation}
  \rho_{A}\sim\frac{J^{1/2}}{t^{1/4}}.
\end{equation}
Note that this will only be valid on times longer than then basic
collision time ($\sim J^{-2}$ for small~$J$) between particles,
with~$\rho_{A}\sim J$ for shorter times.

Thus we see that the convergence to a unique steady state is a slow
power law---slower than in mean field---with the number of `defects'
decaying as~$t^{1/4}$.  Similar arguments should obtain, as long as
the number of transverse dimensions,~$d-1$, is less than~4, yielding a
defect density decaying as~$t^{-(d-1)/4}$ for~$d<5$, the mean field
result~$t^{-1}$ obtaining for~$d>5$, since in high dimensions random
walkers are essentially as likely to annihilate with walkers that
started far away as with those that started nearby. For the
reaction diffusion system, these results have been proven%
.\cite{bramson:asymlett,bramson:spatialabp}

\subsubsection{Inter-row correlations}

We have shown that the channel network should be unique far from the
top of the system. The correlations within one row for the one-deep
model are trivial---just exclusion to no more than one particle on the
inlets. But there will be correlations between one row and another or
between different times. Since in the steady-state each row of
particles will follow the same pattern down the system, the
correlations in occupation probabilities will be time independent and
those between different rows can be obtained by following the
evolution of a single row of particles down the system. We will focus
on the correlation function~$g(x,y)$, defined as the probability that
an outlet at position~$(x',y')$ and one at~$(x'+x,y'+y)$ are both
occupied. Translational invariance of the steady state implies that
this is a function of~$x$ and~$y$ only.

In order to follow the correlations of a row of particles down the
system we use a trick which takes advantage of the fact that for this
row the direction of the primary outlet at each site is only probed
once.  This allows a simple transformation to be made.

At each site we have two equally likely possibilities, that the left
outlet is primary or that the right outlet is primary. If one particle
enters the site then it leaves via the primary outlet which is equally
likely to be to the left or right. If zero or two particles enter the
site then the properties of the site are irrelevant and the same
number of particles pass through both outlets.  The probabilities of
the various outcomes are exactly preserved if instead of choosing
between primary-left and primary-right states we choose between
`crossed' and `uncrossed' configurations.  Any particle now follows
the continuous route ahead of it. A single particle entering (from
either side) is still equally likely to leave to the left or the
right. If zero or two particles enter the situation is also unchanged.
If two particles enter, then we have, essentially, attached labels to
them by distinguishing between the crossed and uncrossed
configurations; these labels are useful in computing correlations but,
of course, they can not appear in physical quantities.

This picture of particles following predetermined crossing routes is
illustrated in Fig.~\ref{fig:ropes}. It shows that the probability
that an outlet passes a particle in one row is just the probability
that the outlet on any specific earlier row on the same route passed a
particle. If the initial condition is that each outlet is
independently occupied with probability~$J/2$, as it is in steady
state, then this condition is trivially preserved for all rows.

Note that this picture is limited to following down a single row of
particles. This can be seen by considering a site in which only one
particle enters, but from the left in one copy and the right in
another. Since these particles should both go out the same primary
outlet, the choice of crossed or uncrossed configurations would need
to be made {\em differently\/} for the two copies.

We have mapped the single row evolution to a collection of
non-interacting particles. The correlation function is thus very
simple in terms of the outlet (or inlet) occupation numbers.  The
correlation function~$g(x,y)$ is determined by whether the two outlets
lie on the same route. If they are on different routes then each has
an {\em independent\/} probability~$J/2$ of holding a particle, so the
probability that both have a particle is~$J^2/4$.  However if the
outlets lie on the same route, they will either both be occupied with
probability~$J/2$ or both be unoccupied.  Thus
\begin{equation}
  g(x,y)=\frac{J}{2}f(x,y)+\frac{J^2}{4}\left[1-f(x,y)\right]
\end{equation}
where~$f(x,y)$ is the probability that the two outlets lie on
the same route.
The truncated correlation function is then simply
\begin{equation}
  g(x,y)-\frac{J^2}{4}=\frac{J}{2}(1-\frac{J}{2})f(x,y),
\end{equation}
the particle-hole symmetry (under~$J\rightarrow2-J$) being manifest.
The function~$f(x,y)$ is just the diffusion kernel on the outlet
lattice, and hence for large~$y$, scales as
\begin{equation}
f(x,y)\approx\sqrt{\frac{2}{\pi y}}\exp(-x^2/2y)
\end{equation}
The simplicity of this correlation function hides some complexity as
this treatment considers only one row of the system. The development
of correlations between different rows is not obtainable by this trick.

Other information on the channel network is readily available,
however. For example, one could ask about the typical vertical
distance between splits and joins as a way to define a correlation
length. Specifically if there is a split (i.e., a doubly occupied
site) at some point~$\vec{s}$ in the network then one can consider
following the primary path from that point down and asking for the
probability of another part of the network joining it as a function of
vertical distance.  This will have two contributions. For short
distances, the dominant processes will be further intersections with
the other route that bifurcated at the initial split. The probability
of this decays with vertical distance as~$y^{-1/2}$, just the
probability of return to the starting point of a random walk in the
transverse direction representing the {\em difference\/} in position
between the two routes.  But in addition there is the chance that
another occupied route will intersect the first. Since the positions
of the other particles in the same row as~$\vec{s}$ is random, for
small~$J$ this extra joining probability will simply be~$J$ at any
distance. Thus the crossover point at which one term dominates the
other is a characteristic vertical length of the network
\begin{equation}
  \xi_{+}\sim\frac{1}{J^2}.
\end{equation}
which can be considered as the typical distance between the `nodes' of
the network; which have at least two completely distinct paths via the
network to either the top or the bottom of the system.

Note, however, that the rejoining probability decays to its long
distance value of~$J$ only as a power law even for
lengths~$\gg\xi_{+}$. There thus appears to be no exponential decay of
correlations in the flowing phase.

For the one deep model an intersection on the channel network has two
incoming and two outgoing channels (for more general models this will
not be the case). The definition of `distinct paths' is that they not
intersect at any {\em sites}. Thus in the `braided' structures
observed in, e.g., the pair of channels on the right side of
Fig.~\ref{fig:nodes}, only the top and bottom intersections of the
braid between the channels are true nodes. By the usual definitions of
percolation theory all of the channel network is part of the
`backbone' while the definition of `nodes' is the same as here.

Near threshold, intersections between channels will be much more
numerous than nodes. The mean distance between intersections along a
channel is~$\text{O}(J^{-1})$, but the characteristic vertical
distance between nodes is~$\text{O}(J^{-2})$ (Eq.~\ref{eq:xiplusj2}).
In mean field theory the distinction between nodes and intersections
is lost and both are spaced by~$\text{O}(J^{-1})$.

Another property of the steady state channel network that follows
immediately from the above picture, is its symmetry: the network will
be statistically up-down symmetric. This is not at all obvious from
the definition of the model, but is a feature of the steady state.
Indeed other features such as the statistics of the `drainage
network'---sites on which an added particle will join the network and
fall to the bottom---are definitely {\em not\/} up-down symmetric This
is illustrated by Fig.~\ref{fig:network} which shows the channel
network and Fig.~\ref{fig:drainage} which also shows the drainage
network and is clearly not up-down symmetric.

\subsection{Beyond the one-deep model}
\label{sec:exten}

We have seen that the simple one-deep model exhibits rather simple
correlations in the steady state channel network, and power law
approach to the steady state. How do these features persist for more
complicated versions of the model?

The two-deep model also has trivial correlations within each row
(which depend on the parameter~$r$), with the outlet occupation
probabilities being independent as in the mean field approximation.
But the inter-row correlations can no longer be treated by the trick
used for the one-deep case. For variants, like the natural three-deep
generalization (which has three parameters analogous to~$r$ of the
two-deep model), the outlet occupation probabilities within one row
are no longer independent. Nevertheless, we conjecture that for the
general synchronous models, the behavior at large scales will be the
same as for the one-deep model, with only short range correlations in
the outlet occupation probabilities within each row, and power law
correlations between rows induced by the diffusion-like behavior of
random walkers.

The convergence to the steady state of the two copies in more general
models is also more complicated than for the one-deep model. As
discussed in the mean field context, the annihilation diffusion
representation is now more complicated with several reacting and
annihilating species. Nevertheless the experience with reaction
diffusion system in other contexts suggests that similar behavior
may persist with~$t^{-1/4}$ decay of the defect density%
.\cite{lee:scalab0,lee:rgka0,peliti:aaa}

\subsubsection{Inter-row Diffusion}

So far we have neglected the possibility of movement of particles
between different rows of particles, i.e., between rows with
different~$y-t$. Each row has been treated independently, every
particle moving at the same speed. A more realistic model might
involve asynchronous dynamics with different particles moving at
different speeds, crudely this can be thought of superimposing some
kind of inter-row diffusion on top of the constant velocity motion.

If the different speeds are a result only of the underlying lattice,
for instance if particles move through some outlets more quickly, then
inter-row motion will be the same in two copies of the system. This
means that the~$A+B\rightarrow\emptyset$ picture of the approach to
equilibrium will survive, with the modification that now the reaction
diffusion occurs in {\em all\/} of the spatial dimensions not just the
transverse ones. The concentrations will then decay as~$t^{-d/4}$
rather than~$t^{-(d-1)/4}$ so that on our two dimensional lattice we
expect decay as~$t^{-1/2}$. If the inter-row motion is slow then there
will be a crossover between the~$t^{-1/4}$ behavior at small scales
and~$t^{-1/2}$ at large scales.

But the inter-row motion might also be affected by the number of
particles at each site. A site with many particles might overflow more
quickly than one with fewer particles. This type of modification is
less easy to accommodate within a reaction diffusion system, but we
expect it will produce qualitatively similar behavior at large scales.

\subsubsection{Thermal Fluctuations}

It is also possible to consider a very simple finite temperature
version of our model within the~$A+B\rightarrow\emptyset$ picture.
Thermal fluctuations would cause a particle to have some chance of
choosing the higher outlet instead of the lower one. This chance would
be stochastic and could happen differently in two copies. If a site
has one particle in each of the two copies and then, due to thermal
fluctuations, moves through different outlets in the two copies then
within the~$A+B$ picture an unlabeled site will split into an~$A$ site
which has a particle in the first copy only, and a~$B$ site which has a
particle only in the second copy. This suggests replacing the reaction
$A+B\rightarrow\emptyset$ with the reversible reaction
\begin{equation}
  A+B\rightleftharpoons\emptyset.
\end{equation}
in this case the statistical steady state will have a finite
concentration of~$A$ and~$B$ sites controlled by the relative rate of
the backward reaction which is determined controlled by the scale of
the thermal fluctuations.

\section{Below Threshold}
\label{sec:below}

We now turn to an investigation of some of the behavior below
threshold. The motion of particles will now only be transient, but one
can also ask about the properties of the stationary state reached when
the transients have decayed.

As discussed in the Introduction, there will be substantial history
dependence. If the force is suddenly increased, many excess particles
will appear and these can collide, resulting in splits and some
particles moving through secondary outlets.

A simpler situation---on which we will focus here---is a slow
adiabatic increase in the force such that at any time there are very
few excess particles and, in the infinitesimally slow limit, there
will be no collisions or splits and every particle will follow a
primary outlet path until it finds a trap. The resulting configuration
will {\em not\/} depend on the order in which the excess particles
appear, thus we can consider starting with all the excess particles in
the system and follow the system down from row to row, letting all the
excess particle enter a site from higher rows, first filling up the
site with any excess then moving {\em all\/} of these out through the
primary outlet.  We can thus `sweep' down the system keeping track of
the number of excess particles on each site which will then be `swept'
into the row below, picking up some extras from the initial excess
particles on the lower row and losing some to unsaturated sites in
that row.

We will specifically consider initial conditions in which each site in
the lattice has an independently chosen number of particles,~$a$,
measured relative to the local capacity.  Negative values of~$a$
represent unsaturated sites (traps) and~$a=0$ represents sites that
cannot trap particles but have no excess particles. The distribution
of values of~$a$ will be represented by~$A_{a}$. It is convenient to
consider the case where~$A_{a}$ only has weight for~$a\geqslant-1$.
Later the particular case
\begin{equation}
  A_{a}=F\delta_{a,1}+(1-F)\delta_{a,-1}
  \label{eq:Adist}
\end{equation}
will be considered in detail,~$F$ representing the applied force.

\subsection{Mean Field Equations}
\label{sec:abovemfe}

In the mean field approximation, the sites on each row are
independent. Thus the distribution of excess particles in the sweep
process is entirely determined by the function~$p_{y}(\tilde{n})$
defined as the probability distribution of the {\em excess\/} number
of particles,~$\tilde{n}\geqslant0$, that are on a site in row~$y$
after the excess particles from higher rows have moved to the site,
but before they have been allowed to move out through the (primary)
outlet.

The probability that a site on the next row down has~$\tilde{n}$
excess particles can be constructed simply by considering the number
of entering particles and the number of particles from the initial
conditions. This equation has a simple form for~$\tilde{n}\geqslant1$
(extra terms are needed for~$\tilde{n}=0$, see the Appendix)
\begin{eqnarray}
  p_{y+1}(\tilde{n})&=&
  c_{0} A_{\tilde{n}}+
  c_{1}\sum_{m=-1}^{\tilde{n}} A_{m} p_{y}(\tilde{n}-m)+ \nonumber\\
  & &c_{2}\sum_{m=-1}^{\tilde{n}} \sum_{\ell=0}^{\tilde{n}-m} A_{m}
  p_{y}(\ell)p_{y}(\tilde{n}-m-\ell).
  \label{eq:btmfrec}
\end{eqnarray}
where the terms represent, respectively, sites with zero, one, or two
inlets that are primary outlets from the row above. The
coefficients~$\{c_{i}\}$ are the probabilities for these numbers of
primary inlets; on the square lattice
\begin{equation}
  c_{0}=\frac{1}{4}\qquad
  c_{1}=\frac{1}{2}\qquad
  c_{2}=\frac{1}{4}.
\end{equation}
The analysis of Eq.~(\ref{eq:btmfrec}) is similar to that carried out
for the continuum problem by~NF and is described in the Appendix.

Far down from the top,~$p_{y}(\tilde{n})$ approaches a limiting
distribution~$p(\tilde{n})$ which can be computed explicitly. For the
specific distribution Eq.~(\ref{eq:Adist}) for the initial excess
particles, it is found that there is a critical value~$F=1/4$ beyond
which no limiting distribution exists. This is the threshold
\begin{equation}
  F_{a}=1/4
\end{equation}
for adiabatic increase of~$F$. Note that this critical force is {\em
  lower\/} than that ($F=1/2$) for the total number of particles to
just fill up all the sites.

At threshold, the limit distribution has a power law tail
\begin{equation}
  p(\tilde{n})\sim\frac{1}{\tilde{n}^{5/2}}.
  \label{eq:pnpower}
\end{equation}
Since at most one excess particle can be trapped by each row going
down the primary outlet path from a site, this power law tail implies
that the probability that some of these particles will fall down by at
least a distance~$y$ before being trapped decays as a power
law---although one that is {\em not\/} simply related to that in
Eq.~(\ref{eq:pnpower}).

For~$F<F_{a}$ the limit distribution decays exponentially, which is
related---although very non-trivially due to the pick up of extra
particles as the sweep continues downwards---to an exponential tail in
the probability that any particle will fall a large distance~$y$
before being trapped. This fall-distance probability can be explicitly
probed by considering the decay, in moving down the system, of a
perturbation in which one extra particle is added; this yields the
probability that the extra particles will move down a distance~$y$
before stopping. From the Appendix it is found that for large~$y$ this
decays as
\begin{equation}
  p_{\text{fall}}(y)\sim \frac{1}{y^{3/2}}e^{-y/\xi_{-}}
\end{equation}
with correlation length~$\xi_{-}$ that
diverges as
\begin{equation}
  \xi_{-}\sim\frac{1}{(F_{a}-F)^{3/2}}
  \label{eq:bmfxi}
\end{equation}
as the threshold is approached, i.e., the critical exponent~$\nu_{a}=3/2$.

After all particles have come to rest the configuration consists (as
discussed in the Introduction and by NF) of saturated trees connected
by primary outlets and terminating (at the bottom) in an unsaturated
terminus site. An excess particle added anywhere on this tree will
fall to the terminus site, possibly thereby extending the tree
downwards if it saturates the terminus site. The correlation length
Eq.~(\ref{eq:bmfxi}) then gives the characteristic vertical length
above which larger trees become exponentially rare. On smaller scales
than~$\xi_{-}$, but near to~$F_{a}$ so that~$\xi_{-}$ is long, the
trees are fractal with number of sites scaling with vertical length
as~$\sim L^{d_{f}}$. In mean field theory, the fractal dimension
\begin{equation}
  d_{f}=\frac{4}{3}.
\end{equation}

{}From calculations analogous to those in the Appendix, and in NF, one
can obtain a scaling form for the probability that a site is on a
saturated cluster of length~$\ell$:
\begin{equation}
  \label{eq:dsfeqn}
  \rho(\ell,F)d\ell\sim\frac{1}{\ell^{\kappa_{a}}}
  \hat{\rho}(\ell/\xi_{-}) \frac{d\ell}{\ell},
\end{equation}
with the scaling function~$\hat{\rho}$ decaying exponentially for
large argument and, in mean field theory,
\begin{equation}
  \kappa_{a}=\frac{2}{3}.
\end{equation}
The probability that an added particle will fall a distance~$y$ before
stopping will have a similar scaling form with the same~$\kappa_{a}$
but a different scaling function.

\subsection{Two dimensions}

In our two dimensional system, the transient and saturated trees that
form on the adiabatic approach to threshold will be similar to those
in the continuum model studied by NF\@. The distribution of tree
lengths will have a form similar to Eq.~(\ref{eq:dsfeqn}) with power
laws out to a vertical distance of order
\begin{equation}
  \xi_{-}\sim\frac{1}{(F_{a}-F)^{\nu_{a}}},
\end{equation}
with the exponents,~$\kappa_{a}$ and~$\nu_{a}$, and the scaling
function~$\hat{\rho}$ different from their mean field values. The
trees will be fractal with some fractal dimension~$d_{f}$. In NF it
was asserted that the width of the trees should scale as the square
root of the length. This may well not be correct and the more general
result
\begin{equation}
  W\sim L^{\alpha_{a}}
\end{equation}
should apply. NF estimated the exponents from numerical simulations
assuming that~$\alpha_{a}=1/2$. Their results~[Eq.~(\ref{eq:nfnum})]
may thus be somewhat off, although the consistency of the analysis
suggests that~$\alpha_{a}$ is probably close to~$1/2$.

The subtle effect, known for other directed percolation-like models,
but missed by NF, is that, although the primary outlet paths are
random walks in the transverse direction, those outlet paths that
wander anomalously far typically will have a different area from which
they can collect excess particles than outlet paths that go closer to
straight downhill. This should make the likelihood of wider trees grow
differently than that of narrower ones resulting in the possibility of
changing the scaling of the large trees and
yielding~$\alpha_{a}\neq1/2$.

In our system, however, there is perhaps some reason to suspect
that~$\alpha_{a}$ could be exactly~$1/2$. Above threshold, the results
on the channel network yield that the exponent,~$\alpha_{+}$, which
relates scaling in the vertical and transverse directions is {\em
  exactly\/}~$1/2$. If there is some form of two-sided scaling, as
suggested in section~\ref{sec:transhist}, then features above
threshold (e.g., finite trees disconnected from the channel network)
would be expected to scale, for intermediate length scales, in similar
ways to those below threshold, i.e., with~$\alpha_{a}$.
How~$\alpha_{a}\neq1/2$ could fit into a two-sided scaling picture
with the channel network with~$\alpha_{+}=1/2$ is unclear; the
simplest scenarios would suggest, therefore, that
$\alpha_{a}=\alpha_{+}=1/2$.

Further numerical simulations in two
dimensions should be able to resolve this, although large sizes and
good statistics will be needed.

In any case, with general~$\alpha_{a}$, the scaling law of NF
relating~$d_{f}$ and~$\kappa_{a}$ in two dimensions should be replaced
by
\begin{equation}
  \label{eq:dsfeqn2}
  d_{f}=1+\alpha_{a}-\kappa_{a}
\end{equation}

\section{Transients and History Dependence}
\label{sec:transhist}

In the previous two sections, we have seen that our model has two
simple limits: below threshold when the force is increased
adiabatically for which no splits occur, and above threshold at constant
current so that the unsaturated sites play no role if the focus is just
on the steady state river network. In both these simple limits, a
crucial part of the physics plays no role, simplifying the behavior
but yielding, not surprisingly, scaling properties for the two phases
which appear to be totally unrelated.

In this section we raise some issues and make a few observations about
the largely unexplored regimes between these two limits. We will focus
on the behavior in an infinite system starting from an initial
condition corresponding to the force to be studied, or
correspondingly, the `sudden' limit where the force is suddenly
increased from zero (i.e., from having no excess particles) to a
particular value. As long as the time to reach steady state in the
bulk is much less than the transit time of particles from top to
bottom, the same behavior should appear also in finite systems except
in a growing boundary region near the top; however in contrast to the
situation studied in Sec.~\ref{sec:ssexact}, the current will be
determined in a non-trivial way from the initial distribution of
excess particle and unsaturated sites.

One might hope to use mean field theory to obtain results for this
situation. Unfortunately the tricks which made mean field theory
tractable for the cases discussed in the previous sections---the
following of individual rows or sets of rows in {\em one\/} pass down
through the system---fail here. This is because whether a particle
gets trapped at a site will depend on the previous history of the
site. Thus the inputs to a site from the row above, while independent
from each other, are correlated with the inputs at earlier times. So
far, we have been unable to derive interesting results from mean field
theory for the sudden limit.

We thus restrict ourselves in this section to qualitative discussions
and general scaling arguments.

\subsection{Below Threshold}

As discussed in the Introduction, the behavior as the force is
suddenly increased is rather different from the adiabatic behavior
discussed in Sec.~\ref{sec:below}. Initially, there will be a large
number of collisions and splits so that some of the particles will
follow secondary outlets and fill unsaturated sites that would not
have been filled with an adiabatic increase in force. We can
characterize the transients and the final stationary state in various
ways. At long times, there will be a small density of particles still
moving and, even though these will tend to be collected together by
their motion downhill, involving mostly primary outlets, the loss of
particles to traps will dominate for~$F$ less than some critical
value~$F_{s}$ ($>F_{a}$), and the last moving particles will follow
paths similar to those in an adiabatic approach. Near~$F_{s}$, we
might expect a power law decay of moving particles at intermediate
times, cutoff by an exponential decay at long times with a
characteristic time scale that diverges at~$F_{s}$.

The stationary state can be characterized in a similar way to the
adiabatic case, by the distribution of saturated trees---connected
by primary outlets---which have the property that an extra particle
which is added on a tree will fall to the (unsaturated) terminus site
before stopping. These trees, as in the adiabatic case, will be
subsets of the complete primary outlet tree which connects all sites.
But they will, in general, be {\em different\/} subsets due to the
splitting processes, and for a given~$F$ will tend to be smaller.
Again, near~$F_{s}$, we expect a distribution of saturated tree
lengths of similar form to Eq.~(\ref{eq:dsfeqn}) but with possibly
different exponents~$\kappa_{s},\nu_{s}$ and scaling of the widths
with the length of clusters of~$W\sim L^{\alpha_{s}}$. The relation
between the number of sites in the large clusters to their length
should again be given by a fractal dimension related to~$\alpha_{s}$
and~$\kappa_{s}$ by Eq.~(\ref{eq:dsfeqn2}).

An outstanding open question is whether or not, in fact, the sudden
approach to threshold is a different universality class from the
adiabatic approach. One might expect this to depend on the rate of
collisions and splits at times of order~$\xi_{-}$ near threshold. This
should also affect whether the relaxation time scales simply
as~$\xi_{-}$ (as in the adiabatic case) or is much longer due to the
splits. We leave these issues for future study.

Another property of the stationary final state is the fraction of
unsaturated sites, which is equal to the number density of saturated
trees (since these include trivial `trees' that consist of only a
single `terminus' site).  This is clearly non-zero below~$F_{s}$, but,
a priori, it could in principle approach zero at threshold if the
exponent~$\kappa_{s}$ is zero so that the large saturated trees
contain most of the sites and are not really fractal,
i.e.,~$L^{d_{f}}\sim L\times L^{\alpha_{s}}$. This question is related
to the issue of unsaturated sites above threshold; the arguments below
strongly suggest that there {\em will\/} be a finite density of
unsaturated sites both at threshold and above.

\subsection{Above Threshold}
\label{sec:transabove}

Above threshold, various aspects of the system started suddenly from
an initial configuration that includes unsaturated sites will be
different from the constant current driven system discussed in
Sec.~\ref{sec:ssexact}.

One set of properties will, however, be the same. At long times, the
system will have a good separation between the moving and stationary
parts of the system and, by the arguments of Sec.~\ref{sec:ssexact},
the channel network will become asymptotically time independent with
the steady state network the {\em same\/} as that for the current
driven case far from the top boundary at the same current density.
(Note that the approach to this network will involve extra splits and
filling in of extra off-network sites resulting in the loss of moving
particles, but the basic property of the moving phase is that these
processes will, at long times, leave a finite fraction of the
particles till moving.)

As mentioned in the Introduction, the dependence of the current
density on~$F$ will be non-trivial, presumably growing as a power
of~$F-F_{s}$ near the threshold as in Eq.~(\ref{eq:currentJ}).

With the system evolving from initial conditions to a steady state
current network, we must right away consider the role of unsaturated
sites. If all the sites eventually become saturated (as in the current
driven case with random time dependent currents inserted at the top),
then the current would be trivially related to the initial particle
density with~$F_{s}$ the point at which the mean excess is zero
and~$J\propto (F-F_{s})$. We will now argue that this will in fact,
{\em not\/} be the case: some fraction of sites will always remain
unsaturated in the infinite-time steady state.

We focus on one site, say the origin, and consider the
probability that it {\em never\/} has a particle pass through it. This
is clearly underestimated by the case with no unsaturated sites and we
thus consider this case; with initial conditions of an excess number
of particles distributed randomly with density~$J$ among the sites in
the one-deep model, so that in the first time step a fraction~$J/2$ of
the outlets carry one particle.

An initial configuration in the row a distance~$h$ above the origin
(i.e., at coordinate~$y=-h$) which has~$h$ consecutive doubly occupied
sites centered at~$x=0$, will ensure that some particle will
definitely reach the origin. But the probability of this extreme event
is~$e^{-ch}$ so that the probability of it not occuring for any~$h$ is
non-zero if~$c$ is large enough. Thus the mechanism which results in
filling in all the sites that are a finite distance from the top in
the case of a random time dependent current inserted at the top, will
{\em not\/} apply here.

But of course, most sites can be reached by less extreme initial
conditions.  We must therefore rely on the convergence properties of
different initial rows passing through the same region as discussed in
Sec.~\ref{sec:ssexact}.

In particular, consider the~$z$ rows starting distances
between~$h$~and~$h+z$ above the origin. As these pass through
height~$h$, there will already be substantial correlations in the
position of the moving particles and ignoring these will tend to
overestimate the probability of reaching the origin; thus in order to
underestimate the probability, we consider
these~$z$~rows as starting with independent random
configurations---$z$~`copies'---at height~$h$.

{}From the aggregation analogy in Sec.~\ref{sec:ssexact}, the simplest
guess is that in each region of width~$\sim\sqrt{h}$ the copy with
most particles in it will essentially `capture' all the particles in
the other copies, so that the positions of the particles in the other
copies at row zero will be subsets of the positions of the dominant
copy in that region. Since the copy with most particles in a region of
width~$\sqrt{h}$ will have
\begin{equation}
  \label{eq:mostnum}
  J\sqrt{h}+\text{O}(h^{1/4}\sqrt{\ln z})
\end{equation}
particles, this suggests that a fraction of order~$h^{-1/4}\sqrt{\ln
  z}$ of row zero that is not on the steady state network will be
reached by some particle from these~$z$~rows.  By choosing,
e.g.,~$z=h$ and considering the set of rows
with~$h=1,2,4,8,\ldots2^{k},\ldots$, we would conclude that the total
fraction~$f_{R}$ of off-network sites reached from any of these sets
of initial rows is
\begin{equation}
  \label{eq:frsum}
  f_{R}<C\sum_{k=1}^{\infty}2^{-k/4}\sqrt{k},
\end{equation}
which is less than one providing~$C$ is sufficiently small; it will be
so for small~$J$. For large~$J$ more care is needed in the estimate of
the effects of nearby rows, but the quantitative mechanism will still
obtain.

This argument is, unfortunately, not quite correct. Because of the
splittings, the majority copy will not, in general `capture' all of
the other copies in a region; and thus Eq.~(\ref{eq:mostnum}) will be
an underestimate of the fraction of sites reached. But for the series
in Eq.~(\ref{eq:frsum}) to converge we need only a much weaker
estimate.  The physics involved in the tendency of the rows from far
above to arrive at the origin looking similar should, we believe, make
a better argument along the above lines possible. But we leave the
necessary work for the future and, at this point, just pose as a
conjecture that {\em a finite fraction of the sites will remain
  unsaturated in the steady state}. Preliminary numerical studies
strongly support this conjecture.

If this conjecture is correct, then various properties of the
off-network sites become interesting. In particular, if an excess
particle is added to the system in the steady state, it will have only
a finite probability~$P_{\infty}$ of moving onto the channel network
and hence down to the bottom of the system. We conjecture that near
threshold the fraction of sites in the `drainage
network',~$P_{\infty}$, behaves as a power law of~$F-F_{s}$ by analogy
with percolation:
\begin{equation}
  \label{eq:pinftperc}
  P_{\infty}\sim(F-F_{s})^{\Gamma_{s}}.
\end{equation}
The statistics of the drainage network on scales smaller than the
distances between nodes on the channel network---$\xi_{+}$
vertically,~$\sim J^{-1}$ in transverse directions---should be related
to the statistics of saturated trees {\em below\/} threshold;
providing a link between above and below threshold scaling. Note
however the subtleties raised at the end of section~\ref{sec:below} if
$\alpha\neq1/2$. We leave analysis of the scaling behavior and the
decay of transients in the moving phase for future work.

\subsection{Adiabatic History}

We close this section with a brief discussion of issues associated
with the alternate, adiabatic history analyzed in
Sec.~\ref{sec:below} below threshold.

If the force is increased at a slow but finite rate, just below the
adiabatic threshold, new excess particles will appear fast enough
that, as they collect on the primary outlet tress, collisions will
become likely resulting in splits. But once the number of particles
becomes sufficiently large above the adiabatic threshold, the moving
particles should be able to establish a steady state channel network
on time scales slower than that characterizing the rate of adding more
excess particles, i.e., the rate of increase of~$F$.

Near~$F_{a}$, there will be non-adiabatic effects, but in the limit of
zero rate of increase of~$F$, it would appear that collisions that are
not present in the steady state channel network will become rare. Thus
we conjecture that in this adiabatic limit, no sites will be filled
except those on the steady state channel network at the final
force~$F>F_{a}$, and those that can be reached by {\em primary\/}
outlet paths (i.e., without splits).

Even if this conjecture is not correct, it appears that the adiabatic
history should yield an alternate steady state above threshold with a
drainage network whose statistics should be related to those of the
saturated trees in the adiabatic history below threshold analyzed in
Sec.~\ref{sec:below}. Again we leave study of these and other
properties, which should be related to those of the continuous fluid
discussed by NF, for the future.

\section{Applications and Conclusions}
\label{sec:conclusions}

We have seen that a very simple model of driven particle motion in a
random system yields very rich behavior, including the development of
distributions of event sizes and fractal spatial structures that are
precursors to steady state flow which occurs, just above a threshold
drive, on a sparse network of channels with most of the system
stationary.

There are various possible applications of at least the qualitative
features of our results.

\subsection{Magnetic Bubbles}

The system of magnetic bubbles in a film driven by a magnetic field
gradient, as studied by Westervelt and collaborators%
,\cite{seshadri:bubrev,hu:collbubb}
 is the simplest
potential application. The model is most directly applicable in the
presence of a high density of strong pinning centers for the
bubbles. The main complication is the effect of the positions of
bubbles in a `trap' on other particles, in particular on how they exit
a trap. If these effects are small, then many of our results should be
directly observable; for example the statistics of the distances
particles move on increasing the drive below threshold, and the
reproducibility and statistics of the channel network above
threshold. Experiments have not yet been performed in this regime but
seem feasible.

For weaker or more dilute pinning, the bubbles will form regions of
hexagonal crystal. On longer scales, however, the strains built up by
the drive through the random medium should be sufficient to break up
the lattice, as pointed out by Coppersmith%
.\cite{coppersmith:instability} %
The sizes of these micro-crystallites can then crudely be thought of
as setting the width of channels; just above the threshold drive, most
will stay still, and motion should occur primarily along a sparse
network of channels. Because of the effect of the bubbles on each
other in this regime the network will probably be more history
dependent than in our models and may evolve somewhat in time as
regions reorient, grain boundaries move, etc. However the qualitative
aspects of broadly distributed effects of increases in the force below
threshold and a sparse channel network above threshold, as well as
history dependence of the threshold force, should persist.

\subsection{Thin Film Superconductors}

The most interesting application is vortex dynamics in thin film
superconductors. But in addition to the effects mentioned above for
magnetic bubbles, there are other complications. If the film is thick,
the vortices will not behave as two-dimensional points, but as lines
bringing in much interesting new physics. If the film is thin, on the
other hand, the interactions between the vortices will be long
range---logarithmic out to the transverse penetration length
($\Lambda_{\perp}=2\lambda^2/d$ with~$d$ the thickness and~$\lambda$
the bulk penetration length), and~$1/r$ at large distances. For
samples small compared to~$\Lambda_{\perp}$, the applied current
density will be relatively uniform and the applied forces on the
vortices roughly equal. For larger films, the current will initially
flow primarily at the edges and only penetrate further concomitantly
with the vortex density building up or being depleted. This effect
should not be too large if the magnetic fields due to the applied
current are small compared to the mean magnetic field, but can play a
role in some regimes. In this regime the vortex density will readjust
as the current is increased to yield roughly constant forces on the
vortices near the critical current. How this will effect the various
features of our model is unclear.

The regime of most direct applicability is a film with holes that in
equilibrium will contain all the vortices. The configuration in a hole
will then only depend on the number of vortices in it, although the
direction in which vortices will be pulled from a hole will in general
depend on the occupation of other holes. Nevertheless, many of the
predictions of the model may be rather directly testable, although
there may be extra history dependence and other modifications due to
the long-range interactions.

The first descriptions of the channel flow of flux lines in thin film
superconductors were numerical simulations performed by Berlinsky,
Brass, Brecht, and Jensen%
.\cite{%
jensen:plaslett,%
jensen:plaslong,%
brass:modquasi,%
jensen:cryoflux} %
Although quantitative comparison with our results is not possible at
this point, their qualitative description of `pulsating flow' along
well-defined channels with the majority of the flux lines remaining
trapped
is the same as ours.
Earlier simulations of this system by Brandt%
\cite{brandt:simtype+brandt:simi+brandt:simii} %
had extensively studied the weak pinning regime where the
Larkin-Ovchinnikov%
\cite{larkin:pinning} %
description of the flux lines as an elastic lattice was appropriate,
but did not investigate how the lattice broke up for higher pinning
strengths.

\acknowledgements

We thank
A.~J. Berlinsky,
E. Domany,
J. Krim,
A.~A. Middleton,
O. Narayan,
S. Redner,
and
M.~S. Tomassone
for useful discussions.
This work was supported by the National Science Foundation via
DMR 9106237 and the Harvard University Materials Research Science and
Engineering Center.

\appendix
\section*{Mean Field Theory Below Threshold }

This mean field calculation for discrete particles has similar
structure to that for continuous systems%
.\cite{narayan:nlflow} %
This similarity contrasts with the situation above threshold where the
discrete and continuous cases are qualitatively different.

The recursion relation for the probability distribution of the excess
number of particles ($\tilde{n}$) at a site in row~$y$,
for~$\tilde{n}\geqslant1$ is~[Eq.~(\ref{eq:btmfrec})]
\begin{eqnarray}
  p_{y+1}(\tilde{n})&=&
  c_{0} A_{\tilde{n}}+
  c_{1}\sum_{m=-1}^{\tilde{n}} A_{m} p_{y}(\tilde{n}-m)+ \nonumber\\
  & &c_{2}\sum_{m=-1}^{\tilde{n}} \sum_{\ell=0}^{\tilde{n}-m} A_{m}
  p_{y}(\ell)p_{y}(\tilde{n}-m-\ell).
  \label{eq:appeqn1}
\end{eqnarray}
For~$\tilde{n}=0$ we have to add the extra terms
\begin{equation}
  c_{0}A_{-1}+c_{1}A_{-1}p_{y}(0)+c_{2}A_{-1}p_{y}^2(0)
\end{equation}
to the right hand side of Eq.~(\ref{eq:appeqn1}), to account for cases
when the site is left below capacity since these also correspond
to~$\tilde{n}=0$.  On the square lattice the probabilities for the
numbers of inlets are
\begin{equation}
  c_{0}=\frac{1}{4}\qquad c_{1}=\frac{1}{2}\qquad c_{2}=\frac{1}{4}.
\end{equation}

Defining transforms
\begin{eqnarray}
  p_{y}(\omega)&=&
  \sum_{\tilde{n}=0}^\infty  p_{y}(\tilde{n})e^{i\omega \tilde{n}} \\
  A(\omega)  &=&\sum_{a=-1}^\infty A_{a} e^{i\omega a}
\end{eqnarray}
the recursion equations become
\begin{eqnarray}
    p_{y+1}(\omega)=
    &\frac{1}{4}&A(\omega)\left[1+p_{y}(\omega)\right]^2+
    \nonumber\\
   &\frac{1}{4}&A_{-1}(1-e^{-i\omega})
   \left(1+\Pi_{y}\right)^2 \label{eq:pwmrec}
\end{eqnarray}
where
\begin{equation}
  \Pi_{y}\equiv p_{y}(n=0)
\end{equation}
gives the fraction of sites in row~$y$ that have no excess particles.

The fixed point condition~$p_{y+1}=p_{y}=p$ has two solutions
for~$p(\omega)$,
\begin{equation}
  \frac{%
    2-A(\omega)\pm
    \sqrt{4[1-A(\omega)]-A_{-1}(1-e^{-i\omega})A(\omega)(1+\Pi)^2}
    }{%
    A(\omega)}
\end{equation}

If we examine the limit~$\omega\rightarrow0$, we see that the argument
of the square root has a root at~$\omega=0$. Such a single root will
give a divergence in~$dp/d\omega$ at~$\omega=0$ which corresponds to
an infinite mean number of particles. This cannot happen below
threshold.  To avoid this the root at~$\omega=0$ must be a double root
which requires
\begin{equation}
\left.\frac{d}{d\omega}\left\{
  4[1-A(\omega)]-A_{-1}(1-e^{-i\omega})
  A(\omega)(1+\Pi)^2
\right\}\right|_{\omega=0}
\end{equation}
to be equal to zero, which means that~$\Pi$ must satisfy
\begin{equation}
  (1+\Pi)^2A_{-1}=-4\bar{a}
\end{equation}
with~$\bar{a}=\sum_{a} a A_{a}$ the mean number of excess particles.
Thus~$\bar{a}$ must be {\em negative\/} up to, and including,
threshold.

Specializing to the case~$A(\omega)=Fe^{i\omega}+(1-F)e^{-i\omega}$
yields the solution
\begin{eqnarray}
p(\omega)&=&
\frac{%
F-1+2e^{i\omega}-Fe^{2i\omega}}{%
1-F+e^{2i\omega}F} \nonumber\\
&-&\frac{%
2(e^{i\omega}-1)
\sqrt{(1-F)(1-2F-2e^{i\omega}F)}}{%
1-F+e^{2i\omega}F}
\end{eqnarray}
As~$\omega\rightarrow0$ this expression has branch cuts at~$F=1$ and
at~$F=1/4$. Identifying the latter as the adiabatic critical
point~$F_{a}=1/4$ we have at criticality,
\begin{equation}
  p_{c}(\omega)=
  \frac{-3+8e^{i\omega}-e^{2i\omega}+2\sqrt{6}(1-e^{i\omega})^{3/2}}{%
    3+2e^{i\omega}}.
\end{equation}
This has a~$\omega^{3/2}$ singularity as~$\omega\rightarrow0$ which
means the critical distribution of excess particle numbers in the
sweep behaves as
\begin{equation}
  p(\tilde{n})\sim\frac{1}{\tilde{n}^{5/2}}
\end{equation}
for large~$\tilde{n}$.

As we approach the threshold distribution from below we must thus
expect the variance of the~$p(\tilde{n})$ distribution to diverge
while the mean remains finite. Calculating explicitly for
\begin{equation}
  f\equiv F_{a}-F,
\end{equation}
we find
\begin{eqnarray}
  \left<\tilde{n}\right>
    &=& 1-2\sqrt{3}f^{1/2}+\text{O}(f^{3/2})\nonumber \\
  \left<\tilde{n}^2\right>-
    \left<\tilde{n}\right>^2
      &\sim& \frac{\sqrt{3}}{4}\frac{1}{f^{1/2}}
\end{eqnarray}
which have the same exponents as the continuous fluid studied by NF.

The steady state mean field calculation gives information about the
distribution of particle occupations. It could be modified as in the
continuous fluid case to provide the distribution of connected cluster
sizes%
.\cite{narayan:nlflow} %
However to obtain information about time
scales (or, equivalently, cluster sizes in the downhill direction) it
is necessary to consider a non-equilibrium calculation.

If the recursion relation between~$p_{y+1}(\omega)$
and~$p_{y}(\omega)$ [Eq.~(\ref{eq:pwmrec})] is considered for
distributions that differ from the fixed point value~$p(\omega)$ by a
small perturbation~$\delta p_{y}(\omega)$ then to linear order in the
perturbation
\begin{eqnarray}
  \delta p_{y+1}(\omega)=
  &\frac{1}{2}&A(\omega)\left[1+p(\omega)\right]\delta p_{y}(\omega)+
  \nonumber\\
  &\frac{1}{2}&A_{-1}\left(1-e^{i\omega}\right)\delta\Pi_{y}
  \label{eq:dprec}
\end{eqnarray}
where~$\delta\Pi_{y}$ is the perturbation of~$p_{y}(x=0)$ away from its
fixed point value.

With a perturbation at an initial row~$y=0$, the recursion relation
is simplified by considering the transforms
\begin{equation}
  \delta p(\omega,z)=\sum_{y=0}^\infty \delta p_{y}(\omega) z^y
\end{equation}
and
\begin{equation}
  \delta\Pi(z)=\sum_{y=0}^\infty \delta\Pi_{y} z^y.
\end{equation}

We can now solve for~$\delta p(\omega,z)$ yielding
\begin{equation}
  \delta p(\omega,z)=\frac{%
    \delta p(\omega,0)+
    \frac{1}{2}zA_{-1}(1-e^{-i\omega})\delta\Pi(z)}{%
    1-\frac{1}{2}zA(\omega)\left[1+p(\omega)\right]}
    \label{eq:dpzsoln}
\end{equation}
where to be self consistent~$\delta p(\omega,0)$ must also be given by
the integral around the unit circle
\begin{equation}
  \delta p(\omega,0)=
  \frac{1}{2\pi i}\oint\frac{\delta p(\omega,z)}{z} dz.
\end{equation}

For the specific case~$A(\omega)=Fe^{i\omega}+(1-F)e^{-i\omega}$,
Eq.~(\ref{eq:dpzsoln}) becomes
\begin{equation}
  \delta p(\omega,z)=\frac{%
    \delta p(\omega,0)+\frac{1}{4}z(1-e^{-i\omega})\delta\Pi(z)}{%
    1-z+z(1-e^{-i\omega})\sqrt{(1-F)[1-2F(1+e^{i\omega})]}}
  \label{eq:dpzsolnspec}
\end{equation}
which has poles at~$z=0$ and at~$z=z(\omega)$ where
\begin{equation}
  z(\omega)=
  \frac{1}{1+(1-e^{-i\omega})\sqrt{(1-F)[1-2F(1+e^{i\omega})]}}
\end{equation}

Performing the contour integral we have the self-consistency
condition
\begin{equation}
  \delta\Pi(z)=\frac{-2\delta p[\omega(z),0]}{%
    zA_{-1}(1-e^{-i\omega{z}})}
  \label{eq:lasttwo}
\end{equation}
where~$\omega(z)$ is the inverse of the~$z(\omega)$ function. We must
find the singular parts of this function which dominate the long time
behavior of the inverse transform of~$\delta\Pi(z)$.

Now~$z(\omega)$ is equal to 1 at~$\omega=0$ and
at~$e^{i\omega}=(1-2F)/2F$. Close to threshold---small~$f$---this
second value is
\begin{equation}
  \omega=-8if+\text{O}(f^2).
\end{equation}
In between these two points on the negative imaginary
axis,~$z(\omega)$ has a simple maximum at
\begin{equation}
  \omega=\omega_{c}=-\frac{16}{3}if+\text{O}(f^2)
\end{equation}
In this vicinity
\begin{equation}
  z=z_{c}+C\left(\frac{\omega-\omega_{c}}{i}\right)^2
\end{equation}
and so~$\omega(z)$ has a branch cut:
\begin{equation}
  \omega=\omega_{c}+i\frac{1}{\sqrt{C}}\sqrt{z-z_{c}}
\end{equation}
where for small~$f$
\begin{equation}
  z_{c}=1+\frac{16}{3}f^{3/2}+\text{O}(f^{5/2})
\end{equation}
and
\begin{equation}
  C=\frac{9}{32}\frac{1}{f^{1/2}}+\text{O}(f^{1/2}).
\end{equation}

This means that for large~$y$ the inverse transform is dominated
by this square root branch cut
\begin{equation}
  \delta\Pi_{y}\sim\frac{f^{1/4}}{y^{3/2}}
  \exp\left[-4y\left(\frac{2f}{3}\right)^{2/3}\right]
  \label{eq:lastone}
\end{equation}
which has the the form
\begin{equation}
  \delta\Pi_{y}\sim\frac{f^{1/4}}{y^{3/2}}e^{-y/\xi_{-}}
\end{equation}
where the vertical length scale,~$\xi_{-}$, diverges at threshold as
\begin{equation}
\xi_{-}\sim f^{-3/2}.
\label{eq:xidiv}
\end{equation}
The~$\nu_{a}=3/2$ correlation length exponent for the cluster size (or
for the response to perturbations) does not correspond to any of those
seen above threshold. It does correspond to that found for the
continuous fluid below threshold by NF.

\begin{figure}
  \caption{Schematic of the lattice
  model.  Circles represent lattice sites and arrows the outlets
  connecting sites.  Thicker lines are primary outlets. The coordinate
  system is also shown.}
  \label{fig:paths}
\end{figure}

\begin{figure}
  \caption{Saturated sites below
  threshold showing the tree structures. Filled circles represent
  saturated sites, unfilled circles represent unsaturated terminus
  sites, and lines show the primary outlets from saturated sites.}
  \label{fig:trees}
\end{figure}

\begin{figure}
  \caption{Steady state channel network
  from an illustrative simulation of the one-deep model with periodic
  boundary conditions.}
  \label{fig:nodes}
\end{figure}

\begin{figure}
  \caption{Route Picture. Lines show
  routes followed by particles. The pattern is specific to a
  particular row of particles moving down through the system. The
  choice of `crossed' or `uncrossed' at each site depends on the
  particle distribution within the row, and is used only for
  computation of single row properties and correlations.}
  \label{fig:ropes}
\end{figure}

\begin{figure}
  \caption{One-deep model above
  threshold, with periodic boundary conditions. Lines show those
  outlets which carry particles in the steady state.  Filled circles
  represent saturated sites, unfilled circles represent unsaturated
  sites. }
  \label{fig:network}
\end{figure}

\begin{figure}
  \caption{Drainage network
  corresponding to the channel network shown in
  Fig.~\protect{\ref{fig:network}}. Thick lines show the channel
  network (outlets through which there is flow in the steady state)
  and thin lines show the drainage network which links sites for which
  an added particle will reach the channel network).}
  \label{fig:drainage}
\end{figure}

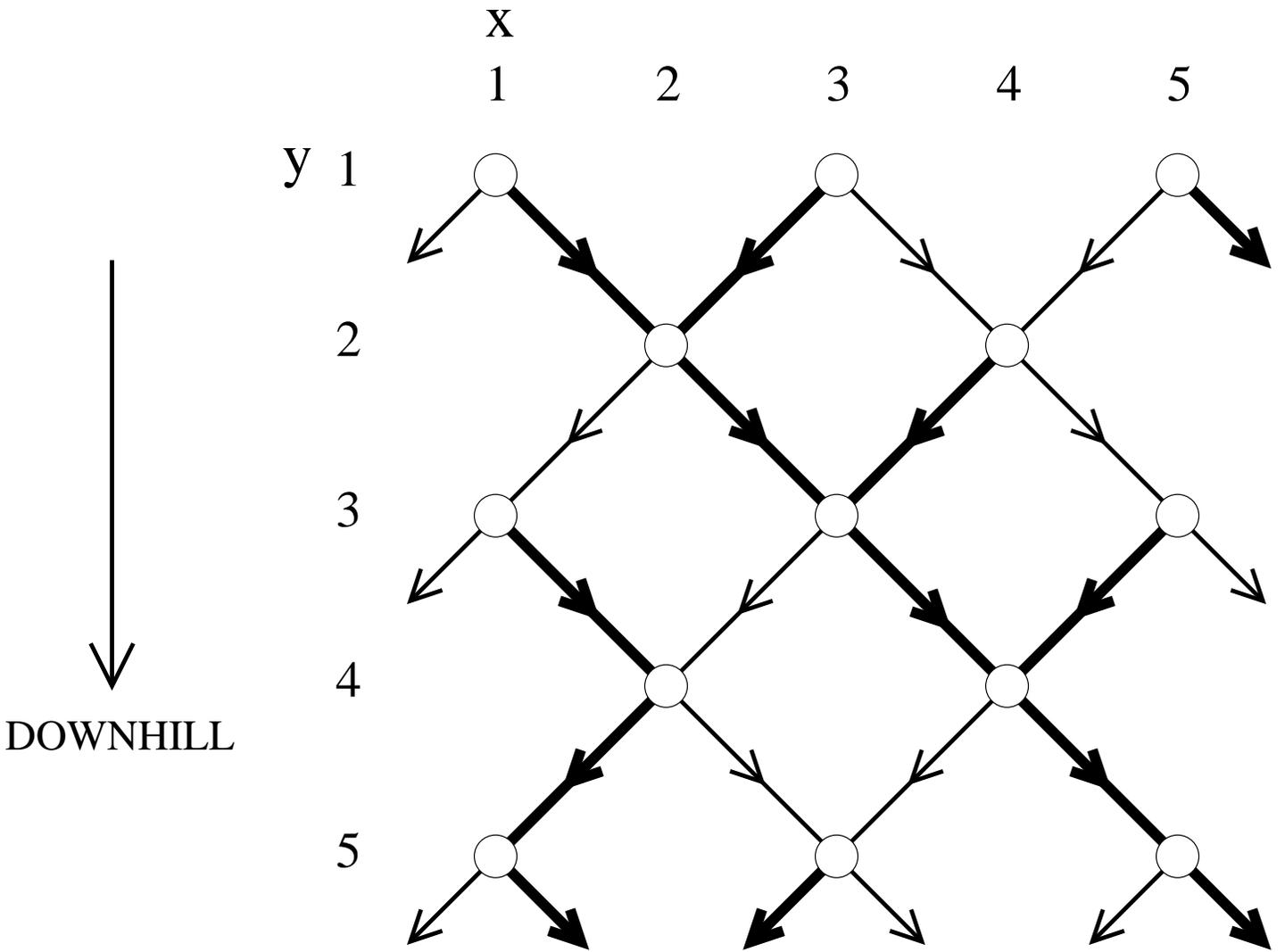
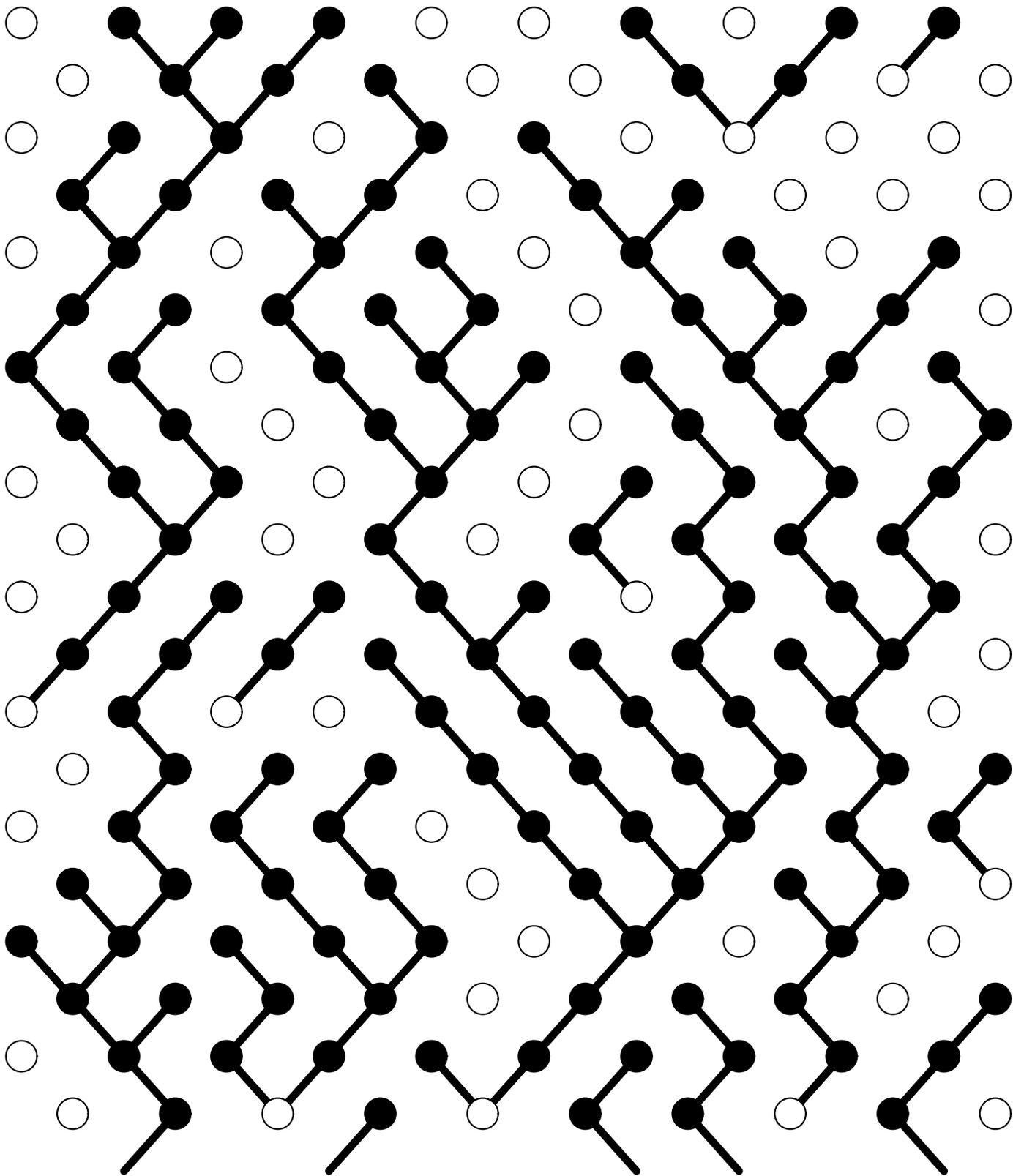
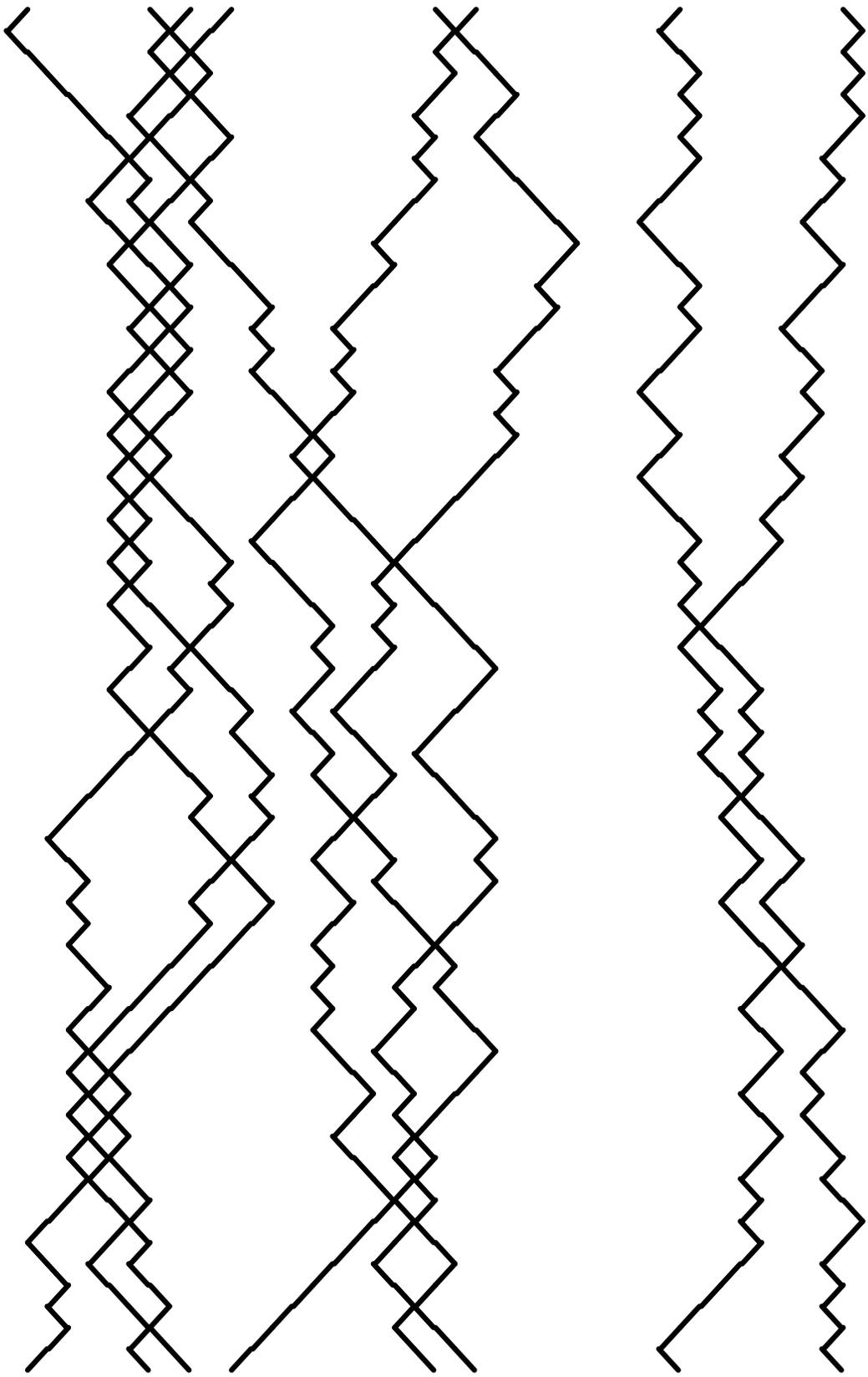
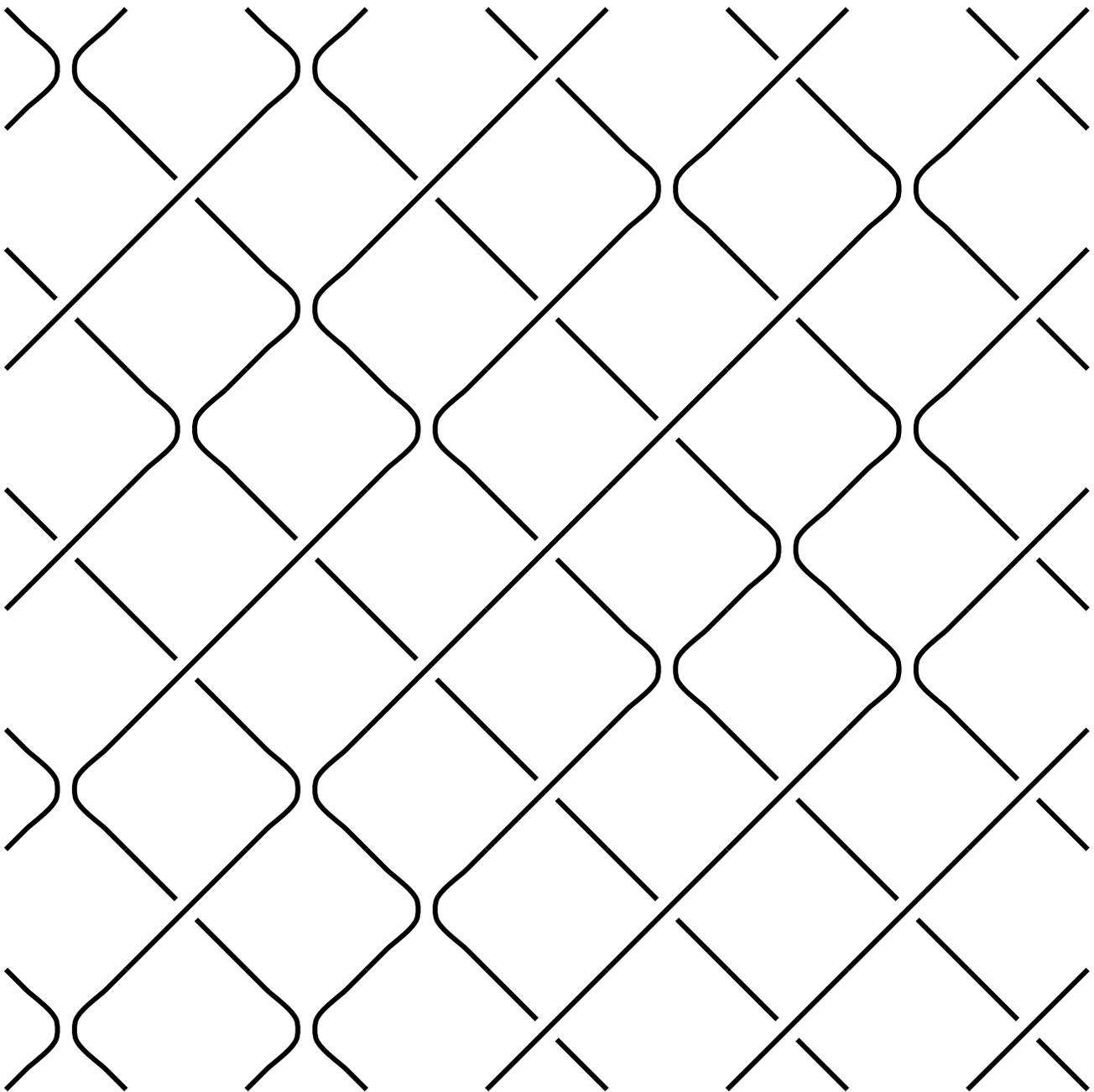
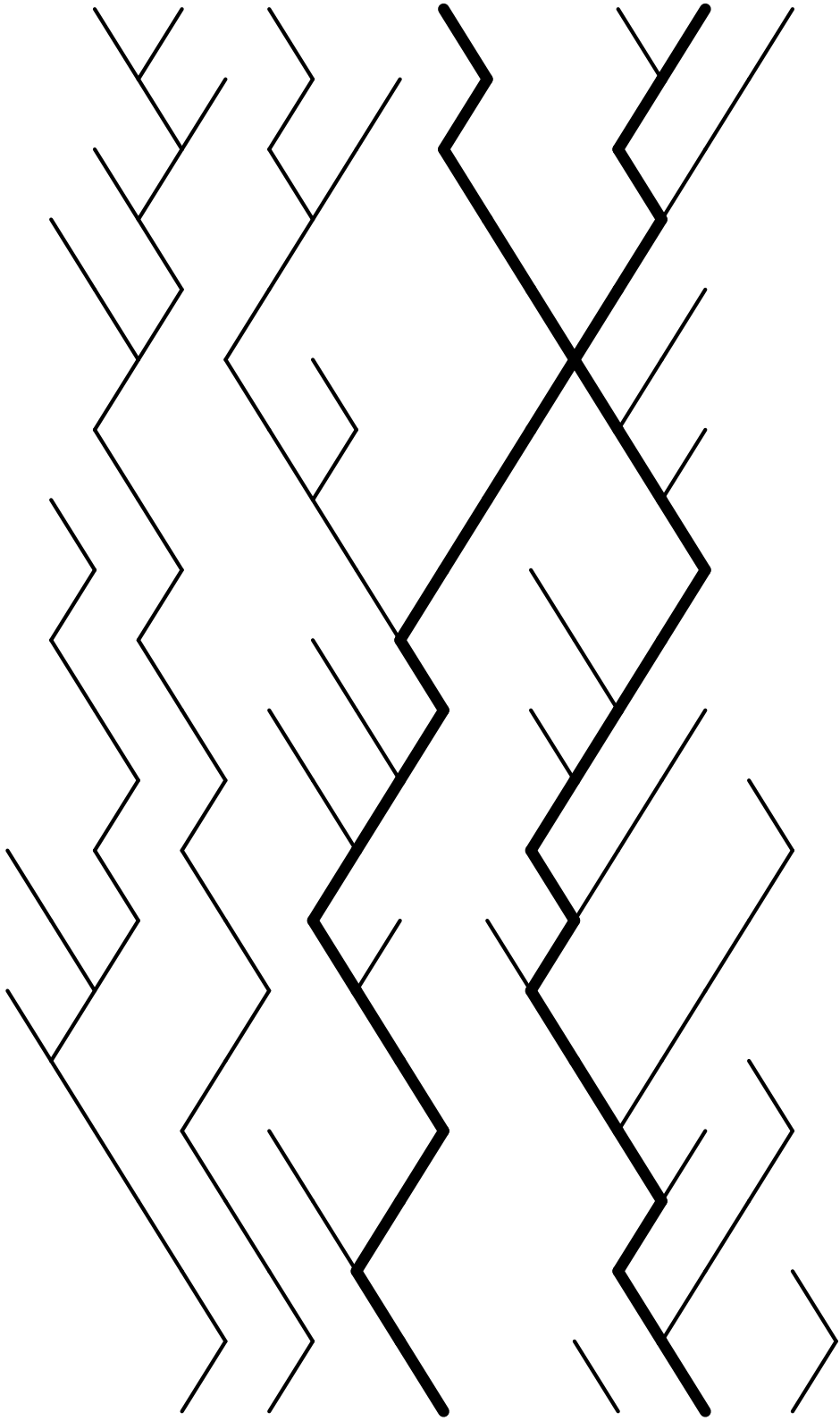
\end{document}